\documentclass[
reprint,
superscriptaddress,
 amsmath,amssymb,
 aps,
prb,
longbibliography,
]{revtex4-2}

\usepackage{graphicx}
\usepackage{dcolumn}
\usepackage{bm}
\usepackage[hidelinks]{hyperref}
\usepackage{color}

\begin{document}

\title{Many-body localization from a one-particle perspective in the disordered 1D Bose-Hubbard model}

\author{Miroslav Hopjan}
\author{Fabian Heidrich-Meisner}
\affiliation{Institute f{\" u}r Theoretische Physik, Friedrich-Hund-Platz 1, 37077 G{\" o}ttingen, Germany}

\begin{abstract}
We numerically investigate 1D Bose-Hubbard chains with onsite disorder by means of exact
diagonalization. 
A primary focus of our work is on characterizing Fock-space localization in this model from the single-particle perspective.
For this purpose, we compute the one-particle
density matrix (OPDM) in many-body eigenstates. We show that the natural orbitals (the
eigenstates of the OPDM) are extended in the ergodic phase and real-space localized when one enters into the
MBL phase. Furthermore, the distributions of occupations of the natural orbitals 
can be used as measures of Fock-space localization in the respective basis.
 Consistent with previous studies, we observe signatures of a transition from the ergodic to the
many-body localized (MBL) regime when increasing the disorder strength.
We further demonstrate that Fock-space localization, albeit weaker, is also evidently present in the distribution of the 
physical densities in the MBL regime, both for soft- and hardcore bosons.
 Moreover, the full distribution of the densities of the physical particles provides a one-particle measure for
the detection of the ergodic-MBL transition which could be directly accessed in experiments with ultra-cold
gases.
\end{abstract}

\maketitle

\section{Introduction}\label{Introduction}

Closed quantum systems with an interplay of interactions and disorder
represent a  paradigmatic case of systems where thermalization is believed to fail
 \cite{Altman15,Nandkishore15,Altman18,Alet18,Abanin19}. 
The original concept of disorder-driven Anderson localization \cite{Anderson65} 
and its generalization to systems of interacting electrons
developed into the more generic framework of many-body localization (MBL) \cite{Gornyi05,Basko06} for closed
quantum systems. The delocalization-to-MBL (or ergodic-MBL) transition 
is an unconventional phase transition at finite energy density, i.e., not related to 
symmetry and not seen in thermodynamics. It is often referred to as
an eigenstate transition \cite{Nandkishore15}. The MBL phase is a state of matter with emergent local 
integrals of motion \cite{Serbyn13b,Huse14,Imbrie16a,Imbrie16b} where
eigenstates exhibit  area-law entanglement \cite{Bauer13,Kjall14,Friesdorf15} and where
slow logarithmic entanglement entropy growth can be observed in global
quenches \cite{Bardarson12,Serbyn13b,Znidaric08}. 
For an overview of this rapidly evolving field, we refer to recent reviews 
\cite{Altman15,Nandkishore15,Altman18,Alet18,Abanin19}.

Insights from numerical investigations of MBL in  spin-$1/2$ XXZ chains (or the equivalent model of spinless fermions)
\cite{Santos04,Oganesyan07,Znidaric08,Pal10,Bardarson12,Kjall14,Luitz14,Friesdorf15,BarLev15,Bera15,Lim16,Khemani16,Khemani17,Alet18} 
by means of exact diagonalization or by means of tensor-network methods
greatly contributed to the current understanding of the MBL phase.
Most of the numerical simulations investigated either the properties of the eigenspectrum
and the eigenstates and the violation of the eigenstate thermalization hypothesis (ETH),
e.g., the level statistics, the number variance, the entanglement entropy, 
and Fock-space localization, or the real-time evolution after a global
quench starting from pure, spatially inhomogeneous initial states.
We note that recently, a controversial discussion emerged on whether the existence of MBL can be 
inferred from finite-size data at all \cite{Suntajs19,Sierant19a,Abanin19a,Panda20}, also questioning the existence of the MBL phase in the
thermodynamic limit \cite{Suntajs19}.
This is related to the intensely discussed question of the exact nature of the
transition (see \cite{Khemani17,Mace20,Goremykina19,Dumitrescu19,Morningstar19,Suntajs20,Laflorencie20}).
These discussions are ongoing, without a final conclusion yet.

Experimental progress has been  made with ultracold atoms \cite{Schreiber15}, trapped ions \cite{Smith16}, and superconducting qubits \cite{Chen17,Roushan17} where
various lattice models with disorder can be emulated. The observation of  signatures of the MBL
phase  was achieved in the quasiperiodic Aubry-Andr\'{e} Fermi-Hubbard model \cite{Schreiber15,Kohlert19},
the disordered Ising model \cite{Smith16}, the disordered Bose-Hubbard model (BHM) \cite{Choi16,Rubio-Abadal19} and  
the quasiperiodic Aubry-Andr\'{e} Bose-Hubbard model \cite{Lukin19,Rispoli19}. 
Recently, the disordered BHM was also realized with interacting photons in an array of superconducting qubits \cite{Chiaro19}.
Most of the experiments carried out with different platforms measure the dynamics of the imbalance decay \cite{Schreiber15,Choi16,Kohlert19} or
the dynamics of the entanglement entropy \cite{Lukin19,Chiaro19}. 

However, so far, only a few numerical studies considered the experimentally relevant disordered BHM  
\cite{Sierant18,Sierant19,Pal19,Orell19,Geisler19,Yao20} (or the BHM with random interactions \cite{Sierant17}).
One reason, perhaps, for the lack of numerical studies are the numerical costs: Full exact diagonalization is feasible only for small system sizes and
the studies are thus limited to 1D \cite{Sierant18,Sierant19,Orell19,Yao20}. For larger 1D or 2D systems, using approximative methods
is unavoidable \cite{Pal19,Geisler19}. Nevertheless, these numerical studies suggest that an MBL phase 
exists in the disordered 1D BHM. The MBL phase was characterized by, for example, the imbalance decay \cite{Sierant18}, the entanglement-entropy growth \cite{Orell19},
the level statistics of many-body eigenspectra \cite{Sierant18,Sierant19,Orell19}, the gap ratio and
the fractal dimension statistics of the full low-energy quasiparticle spectra \cite{Geisler19},
or by the entanglement entropy \cite{Pal19,Orell19}. Furthermore, the existence of one (inverted) \cite{Sierant18,Pal19,Yao20} or more
many-body mobility edges \cite{Geisler19} was proposed. 
Several studies suggest that the existence of double and higher  local occupancies may
favor localization \cite{Michal16,Bertoli18,Bertoli19}, even in the absence of disorder \cite{DeRoeck14,Alex18}.
The understanding of MBL in the disordered BHM is, however,  still far from complete.

Motivated by all these considerations, we here follow an approach based on the one-particle density matrix (OPDM)
computed in many-body eigenstates \cite{Bera15}. By diagonalising the OPDM,   one obtains the natural orbitals and their occupations which can be used
to characterize the real-space localization and Fock-space localization, respectively. This has previously been introduced for spinless fermions in \cite{Bera15} and has been studied in \cite{Bera17,Lezama17,Sheng-Hsuan18,Buijsman18,Villalonga19,Chen20}.
As a main result of this analysis, a  steplike discontinuity in the disorder-averaged occupations of the natural orbitals
was observed, a consequence of Fock-space localization \cite{Basko06,Luitz14,Roy19,Logan19}.
The ergodic phase, by contrast, exhibits a smooth OPDM occupation function, consistent with thermal behavior \cite{Bera15,Bera17}.

Here, we extend these ideas to the bosonic case. In particular, we aim at elucidating the connection between  Fock-space and 
real-space localization in the BHM from the one-particle perspective.
We first revisit the spin-1/2 Heisenberg model, which is equivalent to a model of hardcore bosons,
and which is, at the same time, a standard model for the study of  MBL.
We demonstrate that by diagonalization of the spin-correlation matrix instead of the fermionic OPDM, we also obtain  
natural orbitals and a set of eigenvalues, the occupations.  
The development of a steplike discontinuity in the disordered-averaged spin-projections and
the disordered-averaged occupations of the natural orbitals is observed, analogously to the fermionic case \cite{Bera15} (see also \cite{Ingis16}).
Furthermore, we define a quantitative Fock-space localization measure from the full distributions of the physical spin-projections and the occupations 
of natural orbitals. This measure, which we dub occupation distance,  quantifies the discrete character of the distributions in the MBL phase, related to
the  proximity of many-body eigenstates to Slater determinants (permanents) for fermions (bosons). The system-size dependence of this measure is different in the ergodic and the MBL phase and
the change in the finite-size dependence occurs close to the transition point estimated from other measures \cite{Luitz14,Mace20,Laflorencie20}. 

In the second part, we focus our investigation on the disordered BHM concentrating on densities relevant for recent experiments \cite{Choi16}. 
We first consider the entanglement entropy to show that the disordered BHM indeed exhibits the ergodic-MBL crossover consistent with previous studies \cite{Sierant18,Sierant19,Orell19}.
Then, we diagonalise the bosonic  OPDM to obtain the natural orbitals and their occupations to characterize the real-space localization 
and Fock-space localization.  First, we observe that the natural orbitals are extended 
in the ergodic phase and real-space localized when one enters into the
MBL phase.  We show that the disorder-averaged occupations of the natural orbitals exhibit a step-like structure.
Furthermore, using  our quantitative measure for the degree of Fock-space localization, the occupation distance,  we extract information about the Fock-space 
localization.
Analogously
to spins, the system-size dependence of the occupation distance is different in the ergodic and in the MBL phase.
 Interestingly, the Fock-space localization is also evident in the distributions of physical densities,
which we analyse in the same way as the distributions of the natural-orbital occupations. 
We argue that this type of analysis of the distribution of physical densities may provide an additional 
means to investigate MBL and the ergodic-MBL transition in quantum-gas experiments.

The plan of the paper is the following. We start with the introduction of the one-particle measures both for the spin-$1/2$
case and for bosons in Sec.$~$\ref{sec:intro}. We apply the one-particle characterization to the 1D spin-1/2 Heisenberg model
in the random magnetic field in  Sec.$~$\ref{sec:Heisenberg}. Then, we apply the one-particle characterization to the
disordered BHM in Sec.$~$\ref{sec:BHM}. We conclude our study in Sec.$~$\ref{sec:Conclusions}.

\section{\label{sec:intro}Model and methods}

We first investigate the 1D spin-$1/2$ Heisenberg model with $L$ sites
\begin{equation}\label{spinsham}
H= \sum_{i=1}^{L} \Biggr[\frac{J}{2}(\hat{S}_{i}^{+}\hat{S}_{i+1}^{-}+{\rm H.c.}) +J \hat{S}_{i}^{z}\hat{S}_{i+1}^{z} + h_{i}\hat{S}_{i}^{z}\Biggr].
\end{equation}
Here, $\hat{S}_{i}^{+}(\hat{S}_{i}^{-})$ is a raising (lowering) spin-$1/2$ operator at site $i $, $\hat{S}_{i}^{z}$  measures the $z$-component of the spin and
$h_{i}$ represents a random local magnetic field 
drawn from a box distribution of width $2W$, i.e., $h_{i}\in[-W,W]$. From now on, all energies are expressed in units of the nearest-neighbour  spin-exchange constant $J$. 

Before we introduce the one-particle measure for spins, we review the one-particle characterization for interacting fermions on a tight-binding chain as
originally introduced in Ref.~\cite{Bera15}. 
By virtue of a Jordan-Wigner transformation, Eq.~\eqref{spinsham} can be rewritten as (up to a constant) 
\begin{equation}
\hat H = \sum_{i=1}^{L} \Biggr[-\frac{J}{2}(\hat{c}_{i}^{\dagger}\hat{c}_{i+1}+{\rm H.c.}) + J \hat{n}_{i}\hat{n}_{i+1} + h_{i}\hat{n}_{i}\Biggr]. 
\end{equation}
where $\hat{c}_{i}^{\dagger}(\hat{c}_{i})$ is a creation (annihilation) operator for a fermion at site $i$ and $\hat n_i=\hat{c}_{i}^{\dagger} \hat{c}_{i}$.

For a  given many-body state, $|\psi_{n}\rangle$, 
we measure the one-particle density matrix
\begin{equation}
\rho_{ij}=\langle\psi_{n} |\hat{c}_{i}^{\dagger}\hat{c}_{j}| \psi_{n}\rangle\,.
\end{equation}
The natural orbitals $|\phi_{\alpha}\rangle$ are obtained by diagonalization of the OPDM
\begin{equation}
\rho |\phi_{\alpha}\rangle=n_{\alpha}|\phi_{\alpha}\rangle.
\end{equation}
The eigenvalues $n_\alpha$ are interpreted as occupations of the natural orbitals which sum up to the total number of particles $\sum_\alpha n_\alpha = N$.
We can introduce an associated density operator $\hat n_\alpha=\hat c^\dagger_\alpha \hat c_\alpha$, where  $\hat  c^\dagger_\alpha$ creates a fermion in the 
natural orbital $|\phi_\alpha\rangle$.
In the MBL phase, the natural orbitals exhibit real-space localization and the occupation spectrum reveals the distinctive Fock-space structure of
the many-body eigenstates \cite{Bera15}. The occupation spectrum has a steplike structure with most eigenvalues close to either one or zero and
a discontinuity, thus resembling the momentum distribution of a Fermi liquid \cite{Bera15,Bera17}. 

We now return to the spin representation as used in  Eq.~\eqref{spinsham}.
First, we introduce the expectation value of the
 $z$-component of the spin at site $i$ defined as $s_{i}=\langle\psi_{n} |\hat{S}_{i}^{z}| \psi_{n}\rangle$ in a many-body eigenstate $ | \psi_{n}\rangle$. 
We will argue that the 
expectation values $s_{i}$ can be used as a measure of both  
real-space and Fock-space localization. 
We now introduce the  spin-correlation matrix
\begin{equation}
S^{\pm}_{ij}=\langle\psi_{n} |\hat{S}_{i}^{+}\hat{S}_{j}^{-}| \psi_{n}\rangle,
\end{equation}
which is the analog of the OPDM for spinless fermions.
Note that the spin-correlation matrix does not transform exactly to the OPDM for spinless
fermions under the Jordan-Wigner transformation. Compared to the fermionic OPDM, it acquires additional phases
from the string operators. However, the spin-correlation matrix still provides similar information 
as the OPDM in the case of fermions as we show in Sec.~\ref{sec:spincorr}.
The spin-correlation matrix 
and the $z$-components are connected via $s_{i}=S^{\pm}_{ii}-\frac{1}{2}$.
 
The spin-correlation matrix is  brought to its diagonal form
\begin{equation}
S^{\pm} |\phi_{\alpha}\rangle=s_{\alpha}|\phi_{\alpha}\rangle,
\end{equation}
where $|\phi_{\alpha}\rangle$ are the associated natural orbitals with $s_{\alpha}$ being the respective eigenvalues, i.e., their occupations.
The eigenvalues $s_{\alpha}$ will be used as a measure for  Fock-space localization 
whereas the natural orbitals $|\phi_{\alpha}\rangle$ will be used as a measure for real-space localization.

We further investigate the 1D Bose-Hubbard model with $L$ sites
\begin{equation}
H= \sum_{i=1}^{L} \Biggr[- \frac{J}{2}( \hat{a}_{i}^{\dagger}\hat{a}_{i+1}^{}+{\rm H.c.})+\frac{U}{2}\hat{n}_{i}(\hat{n}_{i}-1)+\epsilon_{i} \hat{n}_{i}\Biggr],
\end{equation}
where $\hat{a}_{i}^{\dagger}(\hat{a}_{i})$ is a creation (annihilation) operator for a boson at site $i$ and $\hat{n}_{i}=\hat{a}_{i}^{\dagger}\hat{a}_{i}$ is the density 
operator at site $i$, $U>0$ accounts for on-site bosonic repulsion and $\epsilon_i$ represents an on-site (diagonal) disorder 
drawn from a box distribution, i.e.,  $\epsilon_{i}\in[-W,W]$. Similarly 
to spins, from now on, all energies are expressed in units of the nearest-neighbour hopping constant $J$.
Note that we use a prefactor of $J/2$ instead of the usual $J$ in front of the hopping term to facilitate the comparison
to the hardcore boson version of the spin Hamiltonian Eq.~\eqref{spinsham}.

For a Bose-Hubbard chain in a given many-body state $|\psi_{n}\rangle$, we measure the set of real-space
site occupations $\{n_i\}$ where the occupation of site $i$ is defined as $n_{i}=\langle\psi_{n} |\hat{n}_{i}^{}| \psi_{n}\rangle$.
Additionally, we
 construct the one-particle density matrix (OPDM) $\rho_{ij}$ defined as
\begin{equation}
\rho_{ij}=\langle\psi_{n} |\hat{a}_{i}^{\dagger}\hat{a}_{j}| \psi_{n}\rangle.
\end{equation}
Note that the OPDM and the site occupancies are connected via $\rho_{ii}=n_{i}$. The natural orbitals $|\phi_{\alpha}\rangle$ 
and their occupations $n_{\alpha}$ are obtained by diagonalization of the OPDM ($\alpha=1,\dots,L)$
\begin{equation}
\rho |\phi_{\alpha}\rangle=n_{\alpha}|\phi_{\alpha}\rangle.
\end{equation}

Note the connection between the spins defined in Eq. (\ref{spinsham}) and the bosons, i.e., the spins can be represented as
hardcore bosons: $\hat{S}^{+}_{i}=\hat{a}^{\dagger}_{i}$, $\hat{S}^{-}_{i}=\hat{a}_{i}$ and $\hat{S}^{z}_{i}=\hat{n}_{i}-1/2$ \cite{Kshetrimayum19}.
The hardcore bosons fulfil the commutation relations
\begin{equation}
[\hat{a}^{\dagger}_{i},\hat{a}^{}_{j}]=[\hat{a}^{\dagger}_{i},\hat{a}^{\dagger}_{j}]=[\hat{a}^{}_{i},\hat{a}^{}_{j}]=0~~~~(i\neq j),
\end{equation}
for different sites and the anti-commutation relations
\begin{equation}
\{\hat{a}^{\dagger}_{i},\hat{a}^{}_{i}\}=1~~~~\{\hat{a}^{\dagger}_{i},\hat{a}^{\dagger}_{i}\}=\{\hat{a}^{}_{i},\hat{a}^{}_{i}\}=0
\end{equation}
for the same site \cite{Rigol09}. Then, the spin-correlation matrix $S^{\pm}_{ij}$ corresponds to the OPDM $\rho_{ij}$ 
in the bosonic picture, i.e., $S^{\pm}_{ij}\rightleftarrows \rho_{ij}$ and $s_{i} \rightleftarrows n_i-1/2$. This also justifies
 the use of the spin-correlation matrix. Therefore, we refer to this object as an OPDM as well.

Apart from  the one-particle measures, we also compute the bipartite entanglement entropy. We split the system into
subsystems A and B, both of size $L/2$, and we expand the eigenstate $|\psi_n\rangle$  as 
$|\psi_n\rangle=\sum_{i}^{}\alpha_{i}|\varphi_{i}\rangle_{\rm A}|\chi_{i}\rangle_{\rm B}$ 
where the $\alpha_{i}$ are positive Schmidt coefficients of the expansion and 
$\{|\varphi_{i}\rangle_{\rm A}\}$ and $\{|\chi_{i}\rangle_{\rm B}\}$ are  orthonormal basis sets in A and B. 
The von-Neumann entropy between the two parts is then defined as the Shannon entropy of the square of the Schmidt coefficients
\begin{equation}
S_{\rm VN}=-\sum_{i} \alpha_{i}^{2} \ln \alpha_{i}^{2}\,. 
\end{equation}

The models introduced above are investigated on systems of
 finite sizes up to $L =18$  (and $10^3$ disorder realizations) for the Heisenberg model and up to $L=14$ (and $10^3$ 
disorder realizations) for the Bose-Hubbard model and periodic boundary conditions are imposed. For spins, the 
overall magnetization is kept to be zero $S^z=\sum_{i}\langle \hat{S}_{i}^{z}\rangle=0$ and for  bosons, we set the filling to $n=N/L= \sum_{i}\langle
 \hat{n}_{i}^{} \rangle/L=0.5$. 

For the spin-1/2 system, we define the target energy density via $\epsilon =\frac{2(E - E_{\rm min})}{E_{\rm max} - E_{\rm min}}$,
where $E$ is the many-body energy of a particular eigenstate and $E_{\rm max}$ and $E_{\rm min}$ are the maximum and minimum 
energy for each disorder realization, respectively. The energy density $\epsilon=1$ corresponds to the middle of the many-body spectrum. 
Full exact diagonalization can be used for system sizes up to $L =16$ (spins) and $L =12$ (bosons), yet we also use the shift-and-invert method
here to reduce the computational effort. 
 For the largest system sizes considered here, $L =18$ (spins) and $L =14$ (bosons), we exclusively use the shift-and-invert method \cite{Pietracaprina18} (without massive
lower-upper decomposition parallelisation). We take the six eigenstates closest to the target energy $\epsilon$ for each disorder realization.
The definition of an energy density for the BHM is more subtle and will be discussed in Sec.~\ref{sec:BHMepsilon}.

\begin{figure*}[t!]
\begin{center}
\includegraphics[width=16cm]{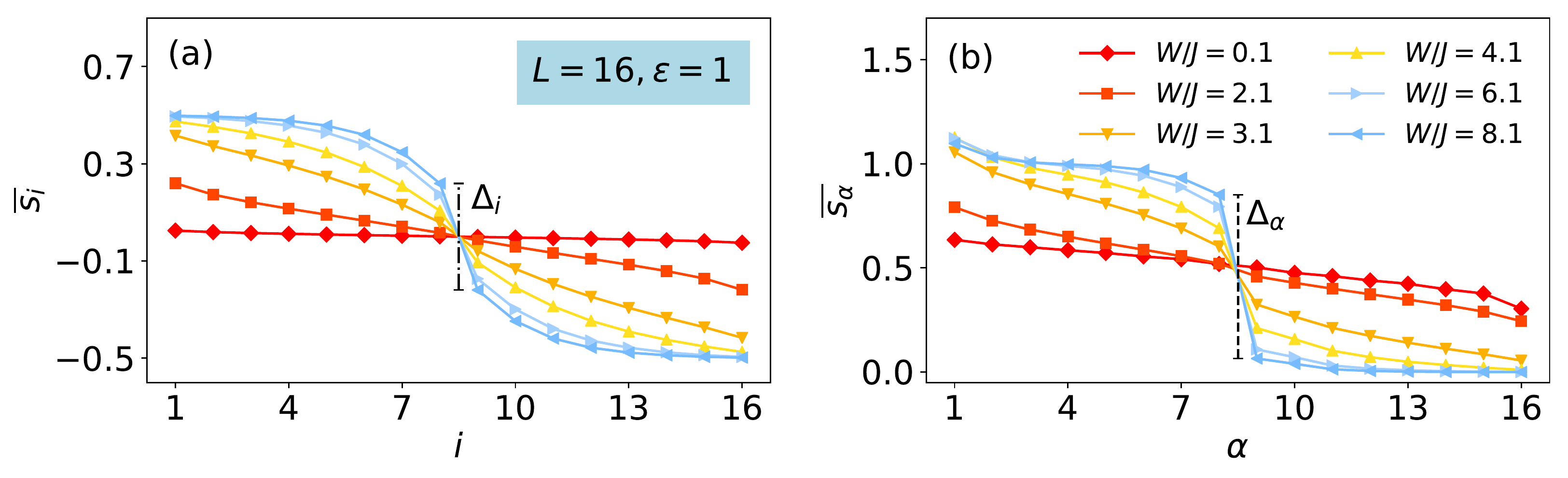}
\caption{{{\it Spin-1/2 Heisenberg chain}: (a) Disorder-averaged spin projection $\overline{ s_i}$ and (b) disorder-averaged OPDM eigenvalues 
$\overline{ s_{\alpha} }$ for $L=16$ and $\epsilon=1$. Both exhibit gaps $\Delta_{i}$ and $\Delta_{\alpha}$ when first ordered
($s_1\geq s_2 \geq \dots \geq s_L$)} and then averaged over disorder realizations.} 
\label{step}
\end{center}
\end{figure*}

\section{\label{sec:Heisenberg} MBL in the 1D Heisenberg model}

\subsection{Disorder-averaged spin projections and OPDM eigenvalues}
\label{sec:spincorr}

We start our discussion with the 1D Heisenberg model. In Fig.$~$\ref{step}, we show the values of the disorder-averaged spin projections $s_i$ and 
$s_\alpha$, which are first re-ordered from the largest value to the smallest one for each eigenstate.  
The disorder average is indicated by the bars. We can clearly observe the development 
of gaps between the values of $\overline{s_{i}}$ and $\overline{s_{\alpha}}$ for $i,\alpha = L/2$ and $i,\alpha= L/2+1$ 
as the disorder strength $W/J$ increases. These gaps are defined as
 $\Delta_{i}=\overline{s_{i=L/2}}-\overline{s_{i=L/2+1}}$ and   $\Delta_{\alpha}=\overline{s_{\alpha=L/2}}-\overline{s_{\alpha=L/2+1}}$.
  Such gaps (or occupation discontinuities)
 were previously reported for spinless fermions \cite{Bera15,Bera17} and for $S=1/2$ spins (and equivalently,  for  hardcore bosons) \cite{Ingis16}.
The gaps reflect the fact that the sites and natural orbitals are either nearly occupied or nearly empty, i.e., the particles
are more real-space localized and the eigenstates are more Fock-space localized. This is a consequence of the existence of emergent
local integrals of motion \cite{Ingis16,Bera17} in the MBL phase. It was also argued that the natural-orbital occupations give a better 
 global approximation to the quasiparticle occupations (i.e., the occupations of the local integral of motions) than 
 the site-occupations or the occupations of Anderson orbitals \cite{Bera17}.
In this respect, the creation operators of natural orbitals are the closest one to the creation operators of quasiparticles (local
 integrals of motions) globally \cite{Bera17}.

\begin{figure}[b!]
\begin{center}
\includegraphics[width=9cm,angle=0]{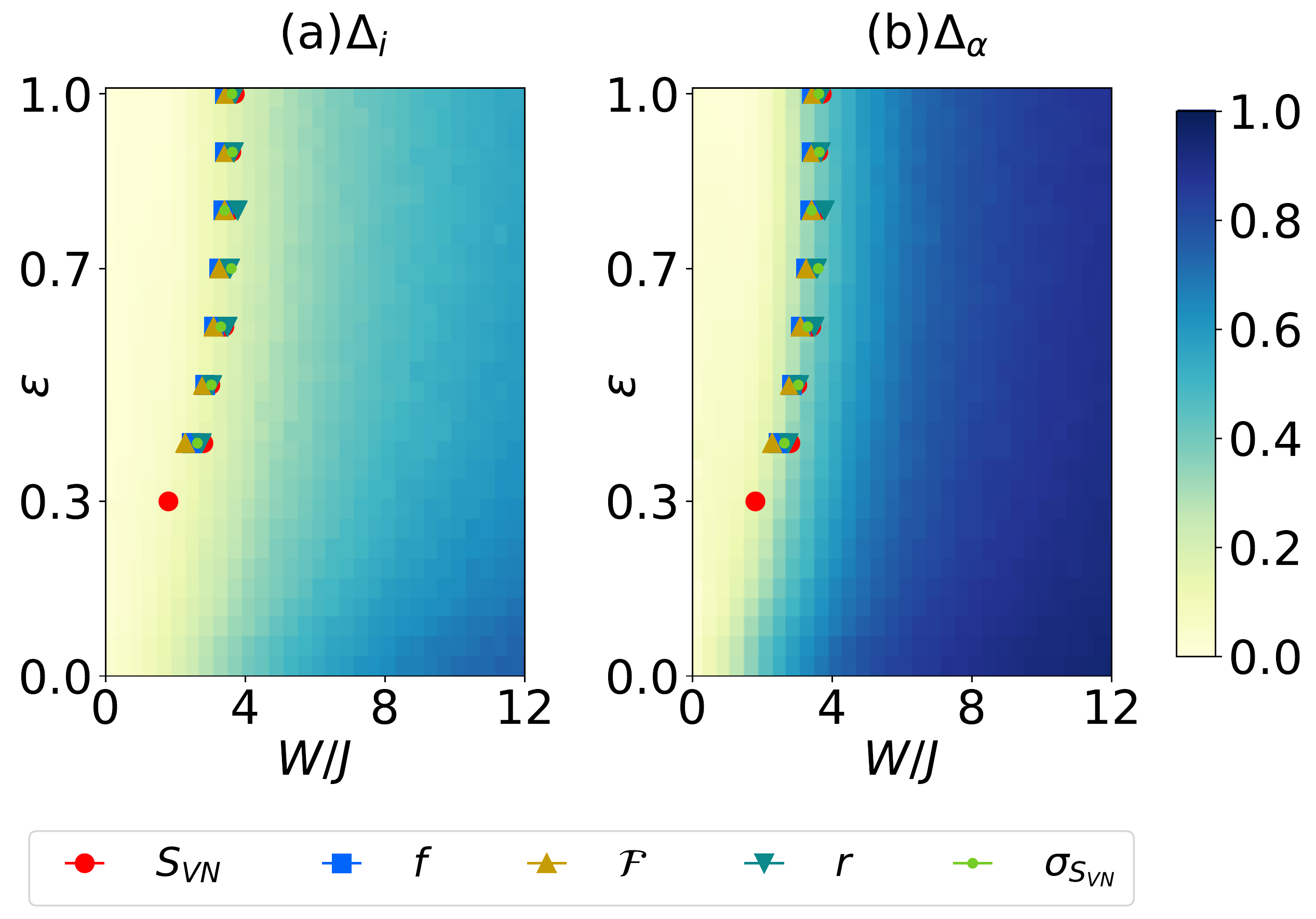}
\caption{{{\it Spin-1/2 Heisenberg chain}:  Dependence of the gaps (a) $\Delta_{i}$ and (b) $\Delta_{\alpha}$ on $W/J$ and $\epsilon$ for $L=16$.
See Figs.$~$\ref{step}(a) and (b) for the definitions of $\Delta_{i}$ and $\Delta_{\alpha}$, respectively. 
The figures include the data from \cite{Luitz14} for the ergodic-to-MBL phase boundary from various measures [$S_{VN}$: an estimate of the boundary between volume
 and area-law scaling of entanglement entropy, $\mathcal F$: bipartite fluctuations of magnetization, $f$: the dynamic fraction, $r$: the ratio of consecutive level spacings,
 $\sigma_{S_{VN}}$: entanglement entropy (fluctuations)].}} 
\label{heatmap}
\end{center}
\end{figure}

In Fig.$~$\ref{heatmap}, we  show  these gaps as a function of disorder strength $W/J$ and energy 
density $\epsilon$ for a fixed system size.  
In the ergodic phase, both  $\Delta_\alpha$ (shown previously in \cite{Bera15,Bera17,Sheng-Hsuan18}) and $\Delta_i$
need to go to zero as $L$ increases, while the occupation discontinuity is expected to persist in the MBL phase, supported by its $L$-dependence as discussed in  \cite{Bera15,Bera17}.
Figures~\ref{heatmap}(a) and (b) also include the numerical results from \cite{Luitz14}
for the transition line between the ergodic and the MBL phase extracted from a number of measures (see the caption of Fig.$~$\ref{heatmap} for details).
According to these data and at energy density $\epsilon=1$, the transition occurs at about $W_c/J\approx 3.6$ \cite{Luitz14}.
This comparison with the behavior of the gaps is rather encouraging.
 The crossover is 
more visible for $\Delta_\alpha$ as the natural-orbital occupations are the superior single-particle measure for  Fock-space localization \cite{Bera17}.
 $\Delta_i$, however,  is the experimentally more accessible quantity 
as it only requires the  measurement of spin projections or densities.  
This motivates our study of distributions of densities for the disordered BHM.

Before moving on, we remark that it is well-known that finite-size data extracted from system sizes $L\leq 26$  can suffer from severe  finite-size effects in the crossover region
\cite{Khemani17,Suntajs19,Sierant19a,Abanin19a,Panda20}. Different quantities exhibit different drifts of transition points (see, e.g., \cite{Kjall14}).
Moreover, there is a range of values reported for the critical disorder strength at, e.g., energy density $\epsilon =1 $ in the literature.
For instance, numerical linked-cluster expansion simulations \cite{Devakul15} or a study of the  imbalance decay in  
Heisenberg chains of $L = 100$ spins \cite{Doggen18} find substantially larger values for the transition point of $W_c/J \approx 4.5$ - $6$.
More recent studies \cite{Chanda20,Laflorencie20} obtain $W_c/J \approx 4.2$ with varying error bars.
Notably, the results of one-parameter scaling ansatzes (see, e.g., \cite{Luitz14}) violate the Harris bound \cite{Harris74,Chandran15,Khemani17}, suggesting that the accessible system sizes may not be in the scaling regime yet.
Some studies propose estimates of how large system sizes need to be to capture the behavior at the transition (see, e.g., \cite{Panda20}).
Even the existence of the MBL phase in the model Eq.~\eqref{spinsham} is discussed controversially \cite{Suntajs19,Sierant19a,Abanin19a,Panda20}. 
The key issue, though, appears to be  that there is no agreement yet on the exact nature of the transition (see, e.g.,  \cite{Khemani17,Mace20,Goremykina19,Dumitrescu19,Morningstar19,Suntajs20,Laflorencie20} for a discussion).

\subsection{Full distributions of spin projections and OPDM eigenvalues}

\begin{figure}[t!]
\begin{center}
\includegraphics[clip,width=8.5cm]{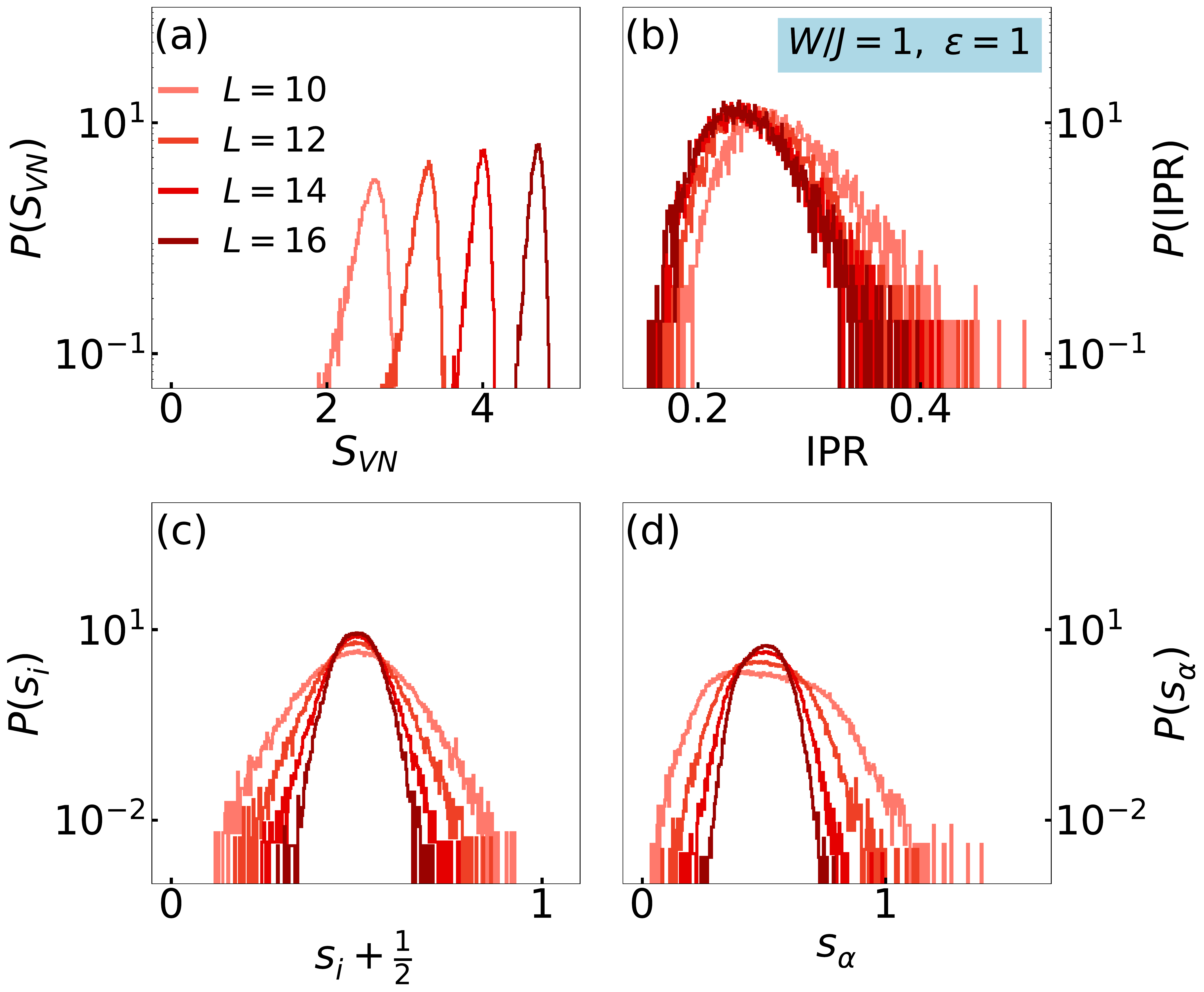}
\caption{{{\it Spin-1/2 Heisenberg chain}: Ergodic phase ($W/J=1, \epsilon=1$).
 Full distributions of (a) von--Neumann entanglement entropy, (b) IPR, (c) spin projections and (d) OPDM occupations 
 for $L=10,12,14,16$.}} 
\label{ergodic}
\end{center}
\end{figure}

\begin{figure}[t!]
\begin{center}
\includegraphics[clip,width=8.5cm]{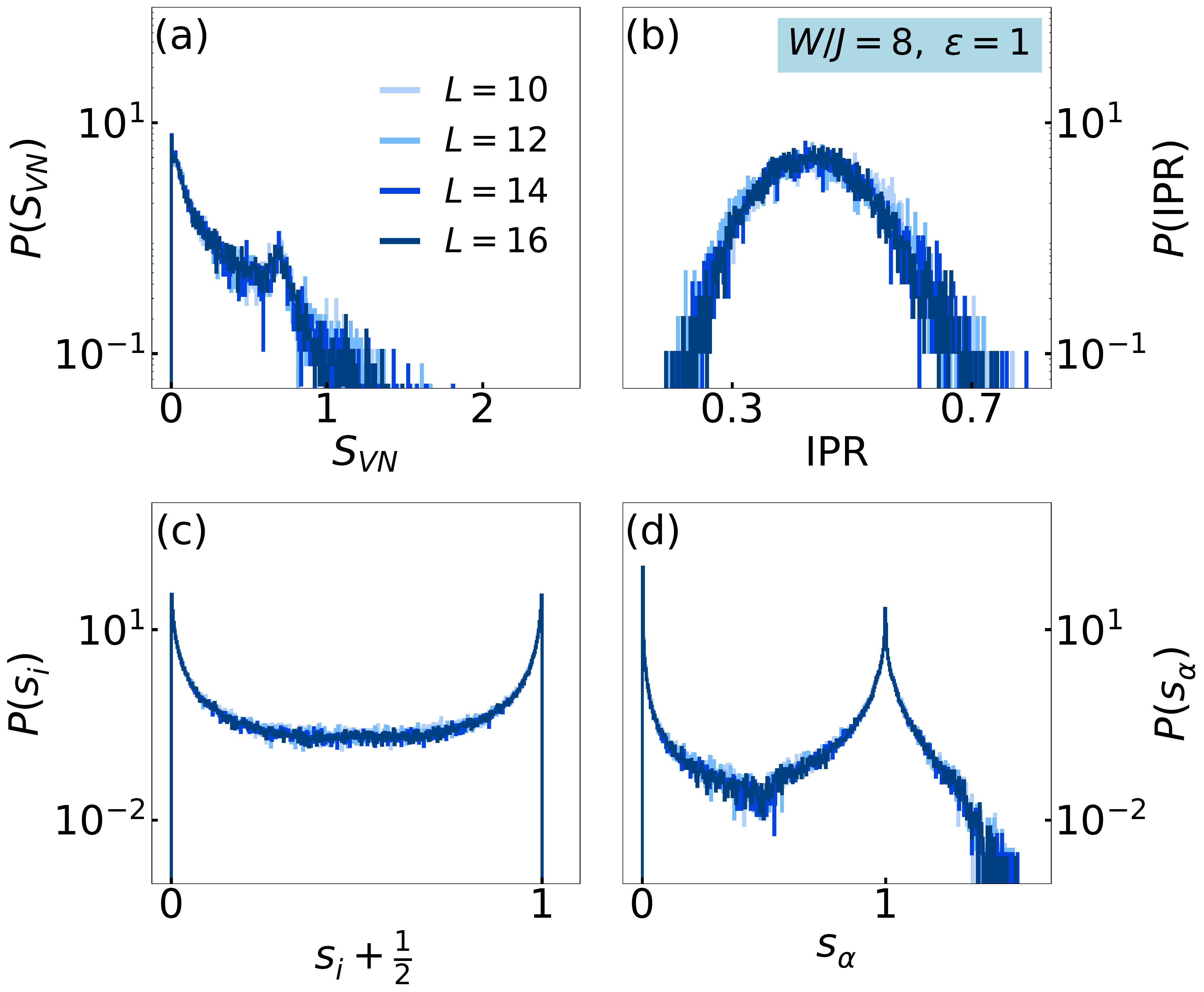}
\caption{{{\it Spin-1/2 Heisenberg chain}: MBL phase ($W/J=8, \epsilon=1$).
 Full distributions of (a) von--Neumann entanglement entropy, (b) IPR, (c) spin projections and (d) OPDM  occupations
 for $L=10,12,14,16$.}} 
\label{localized}
\end{center}
\end{figure}

To better illustrate the behavior of the one-particle observables, it is instructive to plot the full distributions of $s_i$ and $s_\alpha$
deep in the ergodic regime [see Figs.$~$\ref{ergodic}(c) and (d)] and deep in the localized regime [see Figs.$~$\ref{localized}(c) and (d)]. At the same time, we also 
show the distributions of the von-Neumann entanglement entropy $S_{VN}$ in Figs.$~$\ref{ergodic}(a) and \ref{localized}(a).
Finally, we define the inverse participation ratio (IPR)
\begin{equation}
{\rm IPR}=\frac{1}{(S^z+L/2) }\sum_{\alpha=1}^{L} s_{\alpha} \sum_{i=1}^{L} |\phi_{\alpha}^{}(i)|^{4}
\end{equation}
 as a localization measure which contains  information about the real-space localization of the 
natural orbitals $\phi_{\alpha}(i)$. This quantity is shown in Figs.$~$\ref{ergodic}(b) and  \ref{localized}(b).

The system-size dependence of the entanglement-entropy distributions 
for spin-$1/2$ chains was considered before \cite{Lim16,Luitz16,Yu16}. On the ergodic side, the
maximum of the distribution shifts with system size towards higher values \cite{Luitz16} [see Fig.$~$\ref{ergodic}(a)]. Close to the transition,
long tails of low entanglement entropy develop \cite{Luitz16} whereas in the MBL phase, the entanglement entropy distribution does 
not change with the system size \cite{Lim16} [see Fig.$~$\ref{localized}(a)]. A similar behavior was found for the $L$-dependence of the IPR.
In the ergodic phase, the maximum of the IPR distribution shifts towards lower values [see Fig.$~$\ref{ergodic}(b)],
while in the localized regime, the IPR distribution does not change with system size [see Fig.$~$\ref{localized}(b)], consistent with the
results for   spinless fermions \cite{Bera15}.

The distribution of the spin projections $s_{i}$ develops a binary peak structure around the minimal ($s_{i}+1/2=0$) and maximal ($s_{i}+1/2=1$)
possible values with increasing disorder strength $W/J$ \cite{Lim16,Kennes16,Luitz16}. For low
disorder, the distribution depends on system size and becomes sharper as $L$ increases. Moreover, $P(s_i)$ is centered around the average spin projection $ \overline{s_i}+1/2=1/2$ [see Fig.$~$\ref{ergodic}(c)].
 For the larger disorder strength, the distribution 
is practically $L$-independent [see Fig.$~$\ref{localized}(c)]. The distribution of the occupations $s_{\alpha}$ show a 
similar $L$-dependence [see Fig.$~$\ref{ergodic}(d)
and Fig.$~$\ref{localized}(d)]. It develops two peaks when the disorder strength is increased and the peaks are located around 
the integer values $s_{\alpha}=\{0,1\}$, reflecting Fock-space localization \cite{Bera15}. We also see that the  OPDM occupations 
can exceed one. This is due to the bosonic character of the spin system, i.e.,  the spins can be mapped to
hardcore bosons and the hardcore bosons do not obey the strict hardcore constraint in the basis of the natural orbitals. Such 
behaviour was reported before \cite{Ingis16}.

\subsection{Quantitative one-particle measure for Fock-space localization}{\label{sec:measure} }

 \begin{figure}[t!]
\begin{center}
\includegraphics[clip,width=8.5cm]{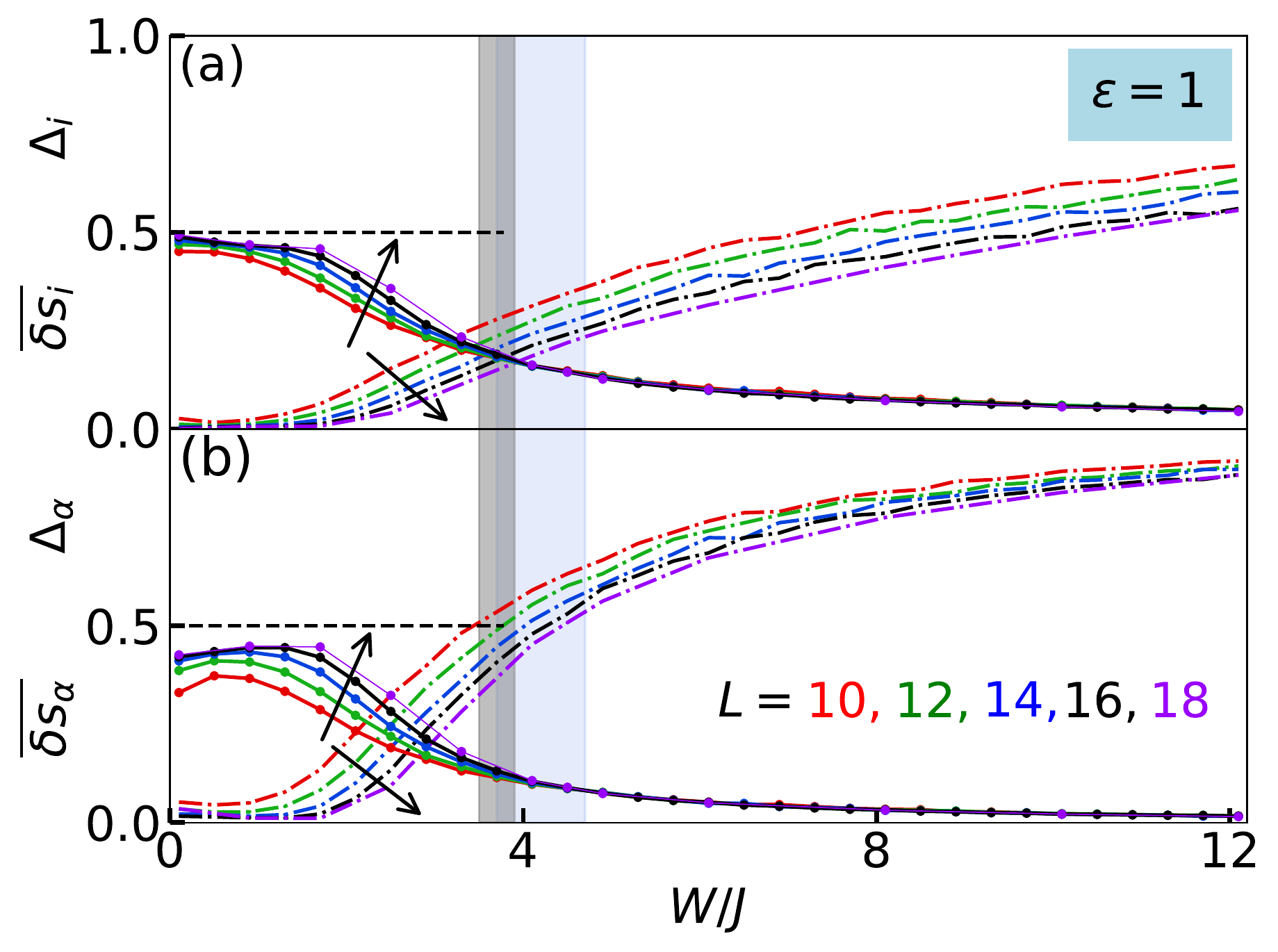}
\caption{{{\it Spin-1/2 Heisenberg chain}: Gaps (a) $\Delta_{i}$ and (b) $\Delta_{\alpha}$ (dashed lines) together with occupation distances, i.e., the
average distances (a) $\overline{\delta s_{i}}$ and (b) $\overline{\delta s_{\alpha}}$ (symbols) to the closest integer
as a function of $W/J$ for $\epsilon=1$. 
See Figs.$~$\ref{step}(a) and (b) for the definitions of $\Delta_{i}$ and $\Delta_{\alpha}$ and
the main text, Sec. \ref{sec:measure}, for the definitions of $\delta s_{i}$ and $\delta s_{\alpha}$, respectively.
The arrows specify increasing system size.
The horizontal dashed line in (a) indicates the filling (of hardcore bosons).  $\overline{\delta s_i}$ is expected to approach this value for $L\rightarrow \infty$ in
the ergodic regime. The horizontal dashed line in (b) indicates an upper bound for $\overline{\delta s_\alpha}$.
For comparison, the vertical lines in grey and blue color mark the position of the ergodic-to-MBL transition estimated from other measures from  Refs.$~$\cite{Luitz14,Mace20}
and Ref.$~$\cite{Laflorencie20}, respectively.}}
\label{scalling}
\end{center}
\end{figure}

We have seen that the distributions $P(s_i)$ and $P(s_\alpha)$ develop peak structures around the integers $s_{i}+1/2=\{0,1\}$ or $s_{\alpha}=\{0,1\}$,
respectively, which reflects  Fock-space localization. In order to
quantify this aspect,  we introduce a measure called {\em occupation distance} computed from each element of the distributions. For the OPDM eigenvalues  $s_{\alpha}$, 
this is defined as
\begin{equation}
\delta s_{\alpha}=\left|s_{\alpha}-[s_{\alpha}] \right|,
\end{equation}
where $[s_{\alpha}]$ is the closest integer to $s_{\alpha}$. For the spin projections of physical particles, we alter the definition to
\begin{equation}
\delta s_{i}= \biggl|s_{i}+\frac{1}{2}-\biggl[s_{i}+\frac{1}{2}\biggr] \biggr|,
\end{equation}
where $[s_{i}+{1}/{2}]$ is the closest integer to $s_i +1/2$. These quantities thus measure the  distance to the closest integers,
or more generally, the distance to the eigenvalues of the corresponding density operators.

 \begin{figure}[t!]
\begin{center}
\includegraphics[clip,width=8cm]{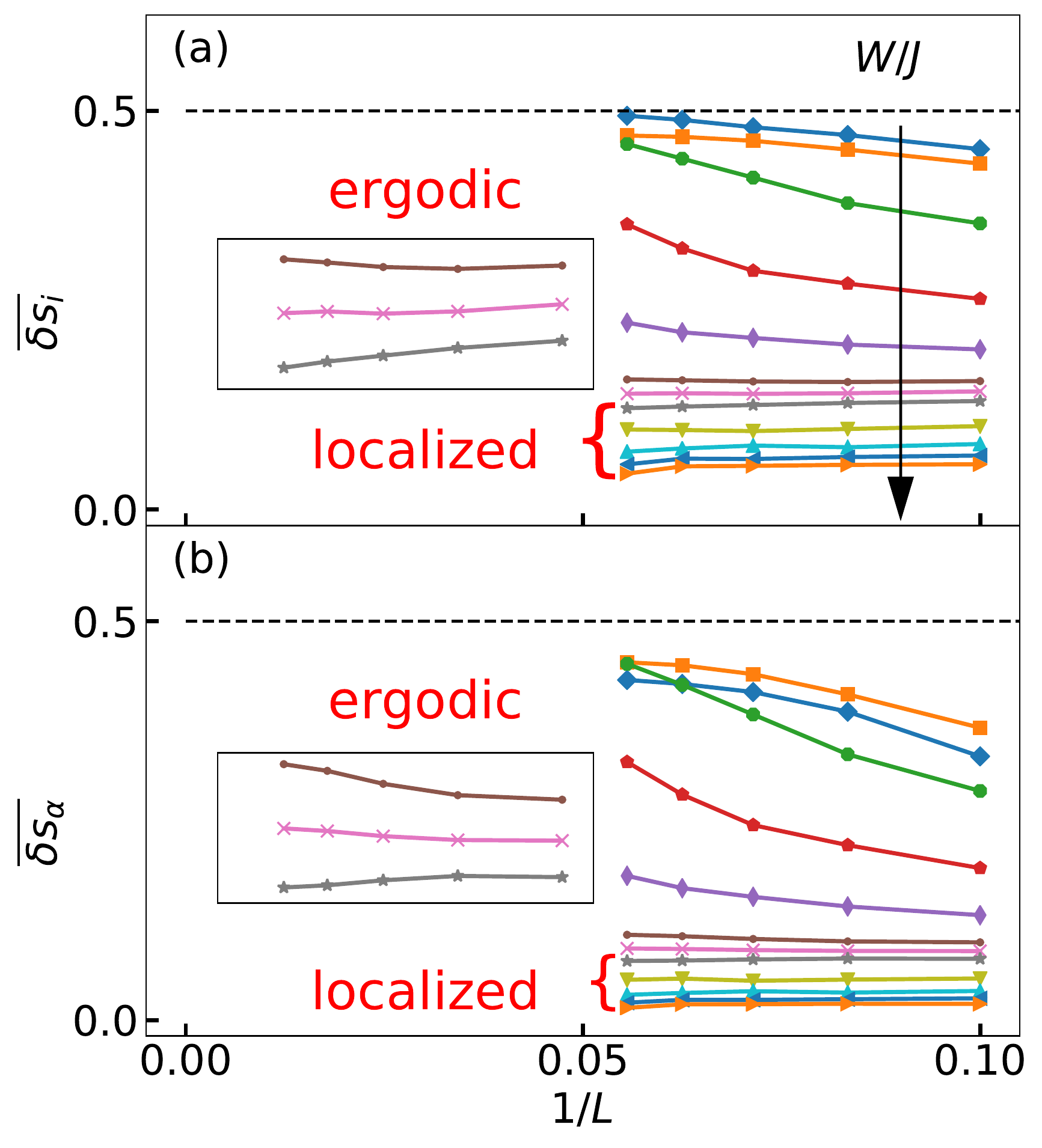}
\caption{{{\it Spin-1/2 Heisenberg chain}:  Average occupation distances (a) $\overline{\delta s_{i}}$ and
(b) $\overline{\delta s_{\alpha}}$  as a function of $1/L$ for $\epsilon=1$ for disorder
strengths $W/J=$ 0.1, 0.9, 1.7, 2.5, 3.3, 4.1, 4.5, 4.9, 6.1, 8.1, 10.1, 12.1 (various symbols). The arrow specifies increasing disorder strength.
The brackets indicate those data sets that we assign to the localized phase.
The insets contain regions zoomed to the data sets for $W/J=$ 4.1, 4.5, 4.9 showing the change of the $L$-dependence trends around
point $W/J\approx4.5$.}}
\label{scalling_new}
\end{center}
\end{figure}

In Fig.$~$\ref{scalling}, we illustrate the dependence of the gaps $\Delta_{i}$ and $\Delta_{\alpha}$  as well as of the disorder-averaged occupation distances
$\overline{\delta s_{i}}$ and $\overline{\delta s_{\alpha}}$ on the disorder strength $W/J$ for the energy density $\epsilon=1$.
We observe that both gaps $\Delta_{i}$ and $\Delta_{\alpha}$ are increasing functions of the disorder strength and that 
the gap $\Delta_{\alpha}$ increases faster than the gap $\Delta_{i}$ which reflects the fact that the basis of natural orbitals
is the better measure for  Fock-space localization. 
When plotted as a function of $1/L$ (not shown here), both gaps extrapolate to a finite value for $W\gtrsim 4J$, 
with $\Delta_\alpha$ extrapolating to larger values than $\Delta_i$ (see Ref.$~$\cite{Bera15} for the $L$-dependence of $\Delta_\alpha$).
Moreover, $\Delta_\alpha$ goes to zero in the ergodic phase as the OPDM occupation function $n_\alpha$ becomes thermal there \cite{Bera15,Bera17}.
It cannot be ruled out that $\Delta_i$ and $\Delta_\alpha$ exhibit a discontinuity at the transition.

 The disorder-averaged distances $\overline{\delta s_{i}}$ and $\overline{\delta s_{\alpha}}$ 
exhibit almost no $L$-dependence for $W/J > 4 $ while for lower disorder strengths,
there is a clear $L$-dependence.  To better observe the change of the behavior, we plot
the $L$-dependences of $\overline{\delta s_{i}}$ and $\overline{\delta s_{\alpha}}$ as a function of $1/L$ in Fig.$~$\ref{scalling_new}. 
At weak disorder, $\overline{\delta s_{i}}$ increases with $L$ and approaches $0.5$ as $L$ increases, as expected for this magnetization sector ($S^z=0$).
Note that a special case of our $\delta s_i$ has recently been studied in \cite{Laflorencie20}. There, specifically $\delta s_i^{\min} = 1/2 - \mbox{max}_{i=1, \dots, L}\lbrace s_i\rbrace$
has been analyzed, which appears to go to zero as $L$ increases in the MBL phase.

A similar increase with $L$ is observed for $\overline{\delta s_{\alpha}}$, where now $0.5$ is an upper bound for $\overline{\delta s_{\alpha}}$.
Since the distribution of $s_\alpha$ is temperature dependent in the ergodic phase \cite{Bera17}, the limit $\overline{\delta s_\alpha} \to 0.5$ is only reached
at exactly infinite temperature.
Note that the limit of $\delta s_i$ and $\delta s_\alpha$ that is approached in
the ergodic phase depends sensitively on the magnetization sector. We will return to this point in the discussion of the
BHM.
For strong disorder,  $\overline{\delta s_{i}}$ and $\overline{\delta s_{\alpha}}$ seem to saturate to values much smaller than $0.5$. 

Remarkably, the point separating these two different  $L$-dependences of $\overline{\delta s_i}$ and $\overline{\delta s_\alpha}$ is close to the estimate of the ergodic-MBL transition point 
extracted from other measures in Ref.$~$\cite{Luitz14} or recently from the multifractal scaling theory discussed in Refs.$~$\cite{Mace20,Laflorencie20}.
The data in the insets of Fig.$~$\ref{scalling_new} show an increase with $L$ for $W/J=4.1$ but a  decrease with $L$ for $W/J=4.9$, while there is  no
 clear $L$-dependence for $W/J=4.5$ suggesting that the change of the behaviour happens somewhere in the interval $W/J\in(4.1,4.9)$.
Thus, there is consistency of our data with those other recent finite-size studies \cite{Mace20,Laflorencie20}
 even though one cannot exclude a drift of the transition point due to finite size-effects \cite{Devakul15,Doggen18,Chanda20,Khemani17,Suntajs19,Sierant19a,Abanin19a,Panda20}.
The results presented above suggest that $\overline{\delta s_{i}}$ and $\overline{\delta s_{\alpha}}$  are useful quantitative measures
for the degree of Fock-space localization (and better suited than $\Delta_i$ and $\Delta_\alpha$)
and motivate us to use analogous measures to study the Fock-space localization in the disordered BHM.

\section{\label{sec:BHM}MBL in the 1D Bose-Hubbard model}

\subsection{Technical aspects and definition of an energy density}
\label{sec:BHMepsilon}

We now turn our discussion to the disordered BHM. Since we consider systems of finite size $L$ with
particle numbers $N=L/2$ and without any hardcore constraint, the local Fock space grows linearly with system size, where 
$M_{\rm loc.}=\{0,1,2,3,\dots,N=L/2\}$. For $L=$ 8, 10 and 12, we construct the Hamiltonian in the full 
many-body basis of size $M=$ 330, 2002 and 12376, respectively \cite{Zhang10,Raventos17}. For $L=14$, we 
perform truncations of the local site occupations in the basis states to 2 and 3 bosons (resulting in 
manageable sizes of the many-body basis of $M=45476$ and $M=69680$, respectively). 

\begin{figure}[t!]
\begin{center}
\hspace{-0.6cm}
\includegraphics[clip,width=8cm]{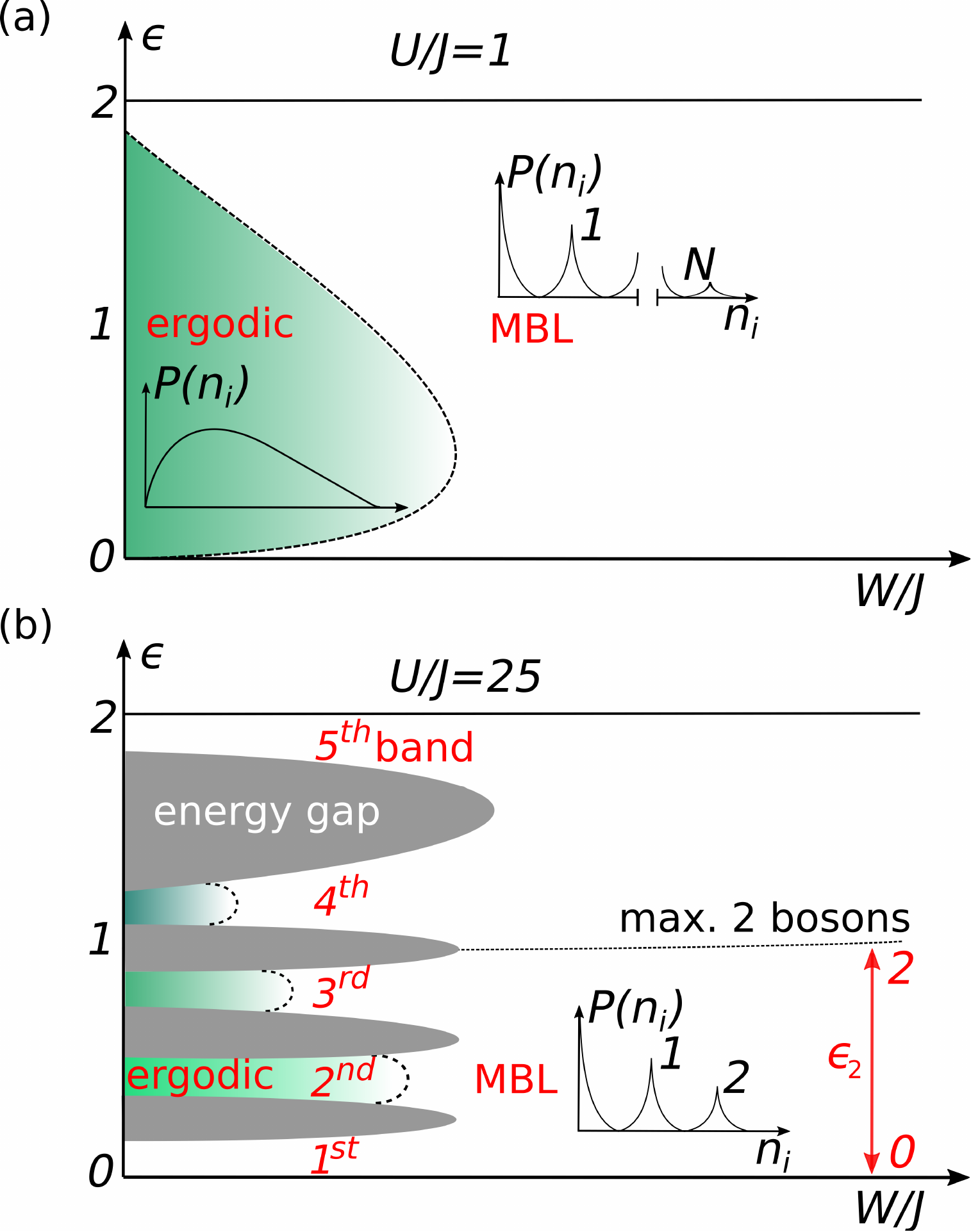}
\caption{{\it Bose-Hubbard model}: Sketch of the many-body eigenspectra in the $\epsilon-W$ plane for (a) $U/J=1$ and (b) $U/J=25$. The sketch corresponds
 to chains of size $L=8$ where, for low disorder, the spectra develop five bands in the
 high-interaction limit $U/J=25$ and where the 3 lowest bands can be characterised by
 the energy density $\epsilon_2$ defined over the sector of eigenstates with maximally 2 bosons per site.} 
\label{sketch}
\end{center}
\end{figure}

In Fig.$~$\ref{sketch}, we show a sketch of the typical eigenspectrum for a system
 in the low-interaction ($U/J=1$) regime [see Fig.$~$\ref{sketch}(a)]
and in the high-interaction ($U/J=25$) regime [see Fig.$~$\ref{sketch}(b)]. 
The large-interaction regime  is more relevant for the
actual experiments \cite{Choi16}.
For the low-interaction regime ($U/J=1$),  the spectrum appears to be continuous. On finite systems, in the high-interaction limit ($U/J=25$),
and for low disorder, the spectrum is divided into well separated bands. The bands are determined
 by the interaction energies of their eigenstates. Typically, the $L$ highest eigenstates in the highest band [see Fig.$~$\ref{sketch}(b)]
correspond to configurations with $N$ bosons occupying mostly one site. By going lower in energy in the many-body spectrum, the
bosons are allowed to be delocalized. The configurations in the lowest bands [see Fig.$~$\ref{sketch}(b)] 
can accommodate typically 1 or 2 bosons per site, respectively. 

For a system of finite size, the many-body spectrum has a maximum energy, which is a function of the total boson number $N$ 
and consequently, the BHM with a fixed filling has an unbounded energy per site in the thermodynamic limit.
In the highest-energy states, all bosons are located mostly at the same site and energies of such states are approximately 
given by $E_{\rm max}\approx UN(N-1)/2$. Then, considering the filling with $N=L/2$, the maximum 
energy can be written as $E_{\rm max}\approx UL(L-2)/8$  and thus
 the maximum energy per site of such states $E_{\rm max}/L\approx U(L-2)/8$ is a linear function of the system size $L$.
 This is different from the case of hardcore bosons where the maximum energy per site is bounded from above. One has to keep this in mind 
 when considering the definition $\epsilon =\frac{2(E - E_{\rm min})}{E_{\rm max} - E_{\rm min}}$ from Sec.$~$\ref{sec:intro} where now
 $\epsilon$ cannot be taken as the energy density.

To obtain a quantity which can be interpreted as an energy density, we look at only the part of the spectrum 
up to a chosen maximal average energy per site. For the system sizes studied here (up to $L=14$), we consider states 
with at  maximum  doubly-occupied sites as such states (for $L=14$, these are the states which have 7 bosons and 3 doubly-occupied sites).
The corresponding energy density $\epsilon_2$ is defined as $\epsilon_{2} =\frac{2(E - E_{\rm min})}{E^{\rm 2}_{\rm max} - E_{\rm min}}$
with respect to the maximum energy of the selected part of the spectrum $E^{\rm 2}_{\rm max}$ [see Fig.$~$\ref{sketch}(b)
for an illustration]. In practice, we first compute the size of the truncated basis $M_{\rm red.}$ by selecting all basis state
which have the local occupancy truncated to $2$. We then construct and diagonalize the Hamiltonian 
in the basis of size $M$ and finally, we compute the energy density $\epsilon_2$ with respect to 
the $M_{\rm red.}$ lowest eigenenergies.

One should note that with an increasing number of sites the number of bands in the $\epsilon_2$ sector  of the
many-body spectra, as defined above, also increases. In the thermodynamic limit, the number of bands will be infinite and the bands will
span the whole range of $\epsilon_2$. However, for the system sizes considered here, the bands remain well
separated for low disorder.  In the following, we focus on the energy density of the second band that roughly corresponds to the middle part of
the $\epsilon_2$ sector, i.e., $\epsilon_2\approx1$ [see Fig.$~$\ref{sketch}(b)] and we discuss the numerical signatures of the ergodic-to-MBL transition there.

\subsection{Entanglement entropy}

 \begin{figure}[t!]
\begin{center}
\includegraphics[clip,width=8.5cm]{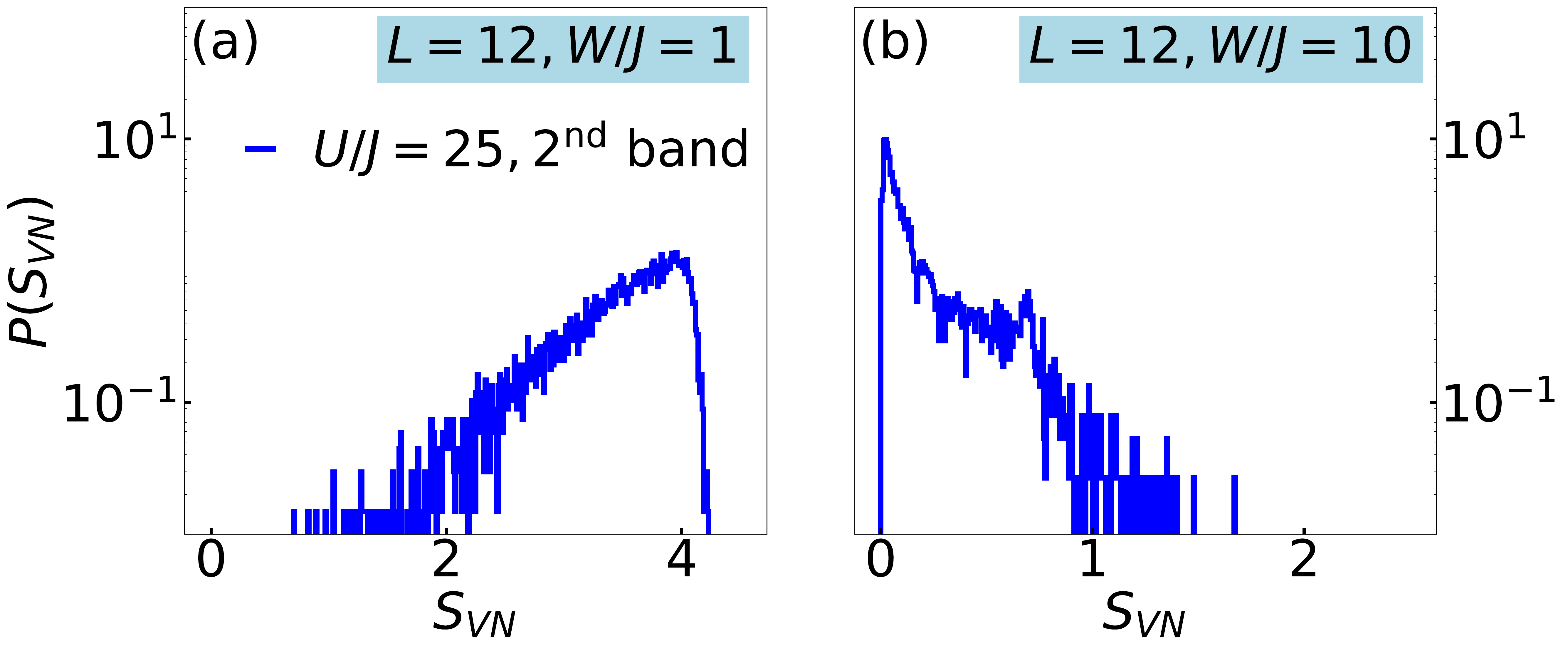}
\caption{{{\it Bose-Hubbard model}: Full distributions of the entanglement entropy of the
 $2^{\rm nd}$ band ($L=12, U/J=25, \epsilon=0.15$) (a) in the ergodic phase and (b) in  the
MBL phase.}} 
\label{entropy}
\end{center}
\end{figure}

The first quantity we look at is the bipartite entanglement entropy as a measure for the 
ergodic-MBL transition \cite{Bauer13}. In Fig.$~$\ref{entropy}, we show representative results for $L=12$  in the second lowest
band for $U/J=25$. For the low disorder $W/J=1$ [see Fig.$~$\ref{entropy}(a)], the entanglement-entropy distributions have a maximum at
a finite value which is the typical shape of this distribution in an ergodic system \cite{Luitz16}. 
For higher disorder [see Fig.$~$\ref{entropy}(b)], the distribution takes the typical shape in the MBL phase with a maximum close to zero 
and a local maximum around $S_{VN}=\ln(2)$ \cite{Lim16,Luitz16}.

\begin{figure}[t!]
\begin{center}
\includegraphics[clip,width=8.5cm]{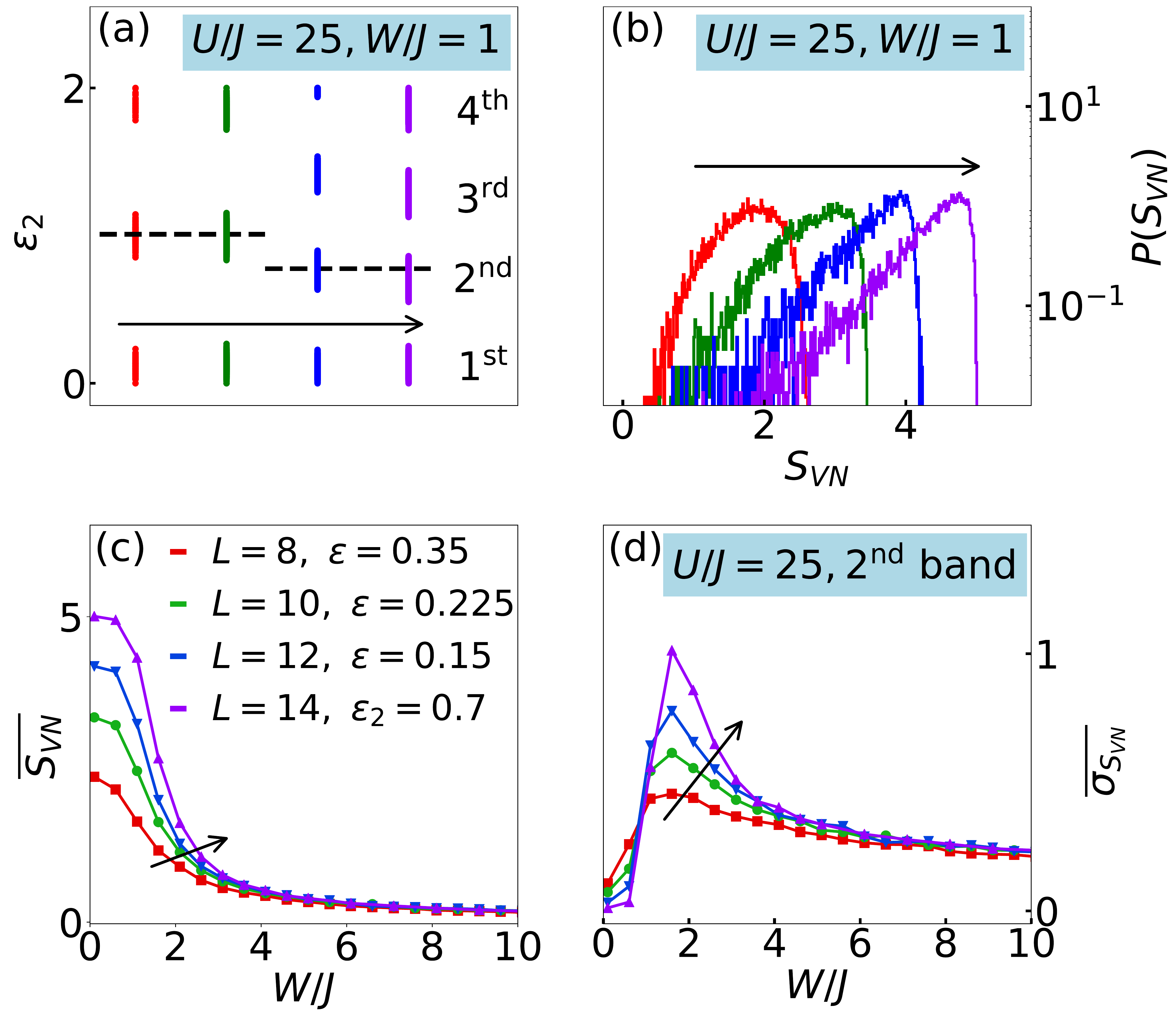}
\caption{{{\it Bose-Hubbard model}: (a) Typical bands of the many-body eigenspectrum expressed in 
the energy density $\epsilon_{2}$ defined over the sector of eigenstates with maximally 
2 bosons per site for system sizes of $L=8,10,12,14$ (with the $L=14$ data from the truncated basis). 
The arrows specify increasing system size.
The dotted lines denote the energy densities of the $2^{\rm nd}$ bands used for the $L$-dependence analysis in (b), (c) and (d). 
In (b), we plot the $L$-dependence  of the full distributions of
the von--Neumann entanglement entropy for the parameters corresponding to the dotted line in (a).
In (c), we plot the $L$-dependence of the average entanglement entropy $\overline{S_{VN}}$ as a function of $W/J$.
In (d), we plot the $L$-dependence of the average fluctuation $\overline{\sigma_{S_{VN}}}$ of the entanglement entropy
 as a function of $W/J$. The arrows specify
increasing system size.
}} 
\label{entropy_scal}
\end{center}
\end{figure}The arrows specify increasing system size.

\begin{figure}[t!]
\begin{center}
\includegraphics[clip,width=8.5cm]{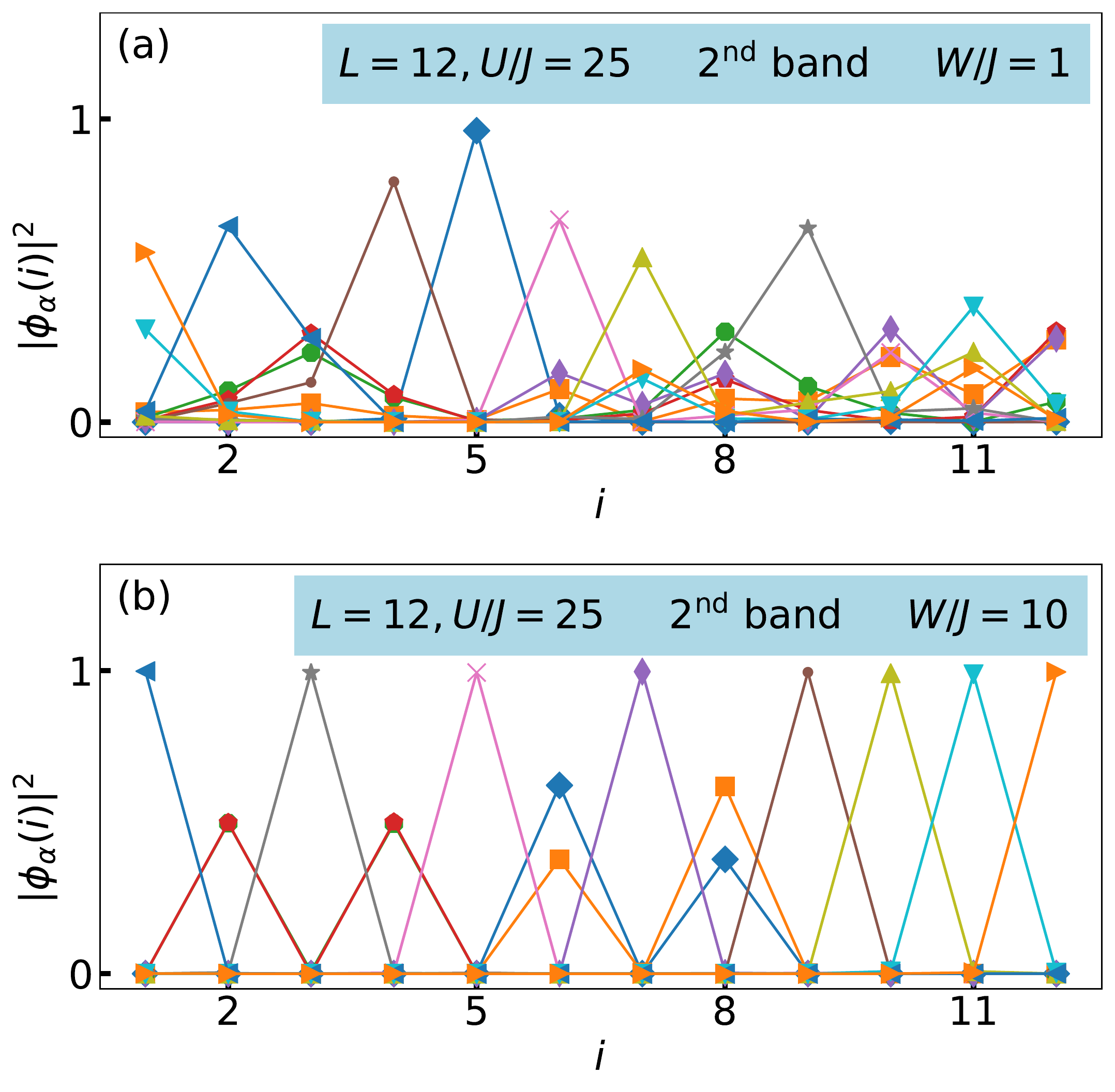}
\caption{{{\it Bose-Hubbard model}: Example of all natural orbitals of the OPDM (various symbols) computed from one randomly chosen eigenstate
of the $2^{\rm nd}$ band ($L=12, U/J=25, \epsilon=0.15$) (a) in the ergodic ($W/J=1$) and (b) in the MBL regime ($W/J=10$). }} 
\label{localizedbos}
\end{center}
\end{figure}

In Fig.$~$\ref{entropy_scal}, we show the $L$-dependence of the entanglement entropy.
 The second lowest bands for $L=$ 8, 10, 12 and 14 have a similar energy density $\epsilon_2$ [see Fig.$~$\ref{entropy_scal}(a)]. 
 For low disorder $W/J=1$, the distributions of the entanglement entropy exhibit a 
shift of the position of their maxima towards higher values [see Fig.$~$\ref{entropy_scal}(b)]. This is the typical $L$-dependence in the ergodic regime \cite{Luitz16}.
At high disorder $W/J=10$, the distribution is $L$-independent (not shown).
In Figs.$~$\ref{entropy_scal}(c) and (d), we plot the average entanglement entropy $\overline{S_{VN}}$ and the average fluctuation $\overline{\sigma_{S_{VN}}}$ of the 
entanglement entropy as a function of the disorder strength $W/J$, respectively. We observe a large and system-size dependent average entropy for
values $W/J \in (0,2.5)$. By contrast,   for values $W/J > 4$, the entropy is close to zero for all considered system sizes. This is also reflected in the fluctuation
of the entanglement entropy which has a maximum value close to $W/J\approx2$. This maximum  shifts to larger values  with increasing system size. The large fluctuations of the entanglement entropy 
are usually interpreted as a numerical signature of the ergodic-MBL transition and they are expected to diverge at the transition for $L\rightarrow\infty$ \cite{Kjall14}.
Thus, from the visual inspection of our finite-size numerical data, we can estimate that the transition happens somewhere at $W_c/J \approx 2$.
By using  the one-parameter scaling ansatz of Refs.$~$\cite{Luitz14,Khemani17}, namely $\overline{S_{VN}}/S^{\rm Page}_{VN}=g[L^{\frac{1}{\nu}}(W-W_c)]$,
where $S^{\rm Page}_{VN}$ is the Page value for a random pure state \cite{Page93},
we  find an estimate for the transition point of $W_c/J = 2.0(1)$. However, similarly to the study of spins in Ref.$~$\cite{Luitz14}, 
the estimate for the exponent $\nu=0.80(5)$ violates the Harris bound \cite{Harris74,Chandran15,Khemani17} 
and one can expect that the true transition point is at a higher value of $W/J$ than the one obtained from the one-parameter scaling estimate. 

\subsection{Natural orbitals and IPR }

In this subsection, we show that the ergodic-MBL transition is also reflected  in properties of the natural orbitals. In Fig.$~$\ref{localizedbos}(a), we plot
all natural orbitals for one randomly chosen eigenstate in the ergodic phase for a low disorder strength ($W/J=1$), while in  Fig.$~$\ref{localizedbos}(b), we
plot all the natural orbitals for one eigenstate in the MBL phase for a high disorder strength ($W/J=10$). For low disorder, the natural orbitals are delocalized 
spanning the whole system [see Fig.$~$\ref{localizedbos}(a)]. On the other hand, from Fig.$~$\ref{localizedbos}(b), a localization of the
natural orbitals by disorder can  clearly be observed, similar to the localization of the natural orbitals for fermionic systems.

\begin{figure}[t!]
\begin{center}
\includegraphics[clip,width=8.5cm]{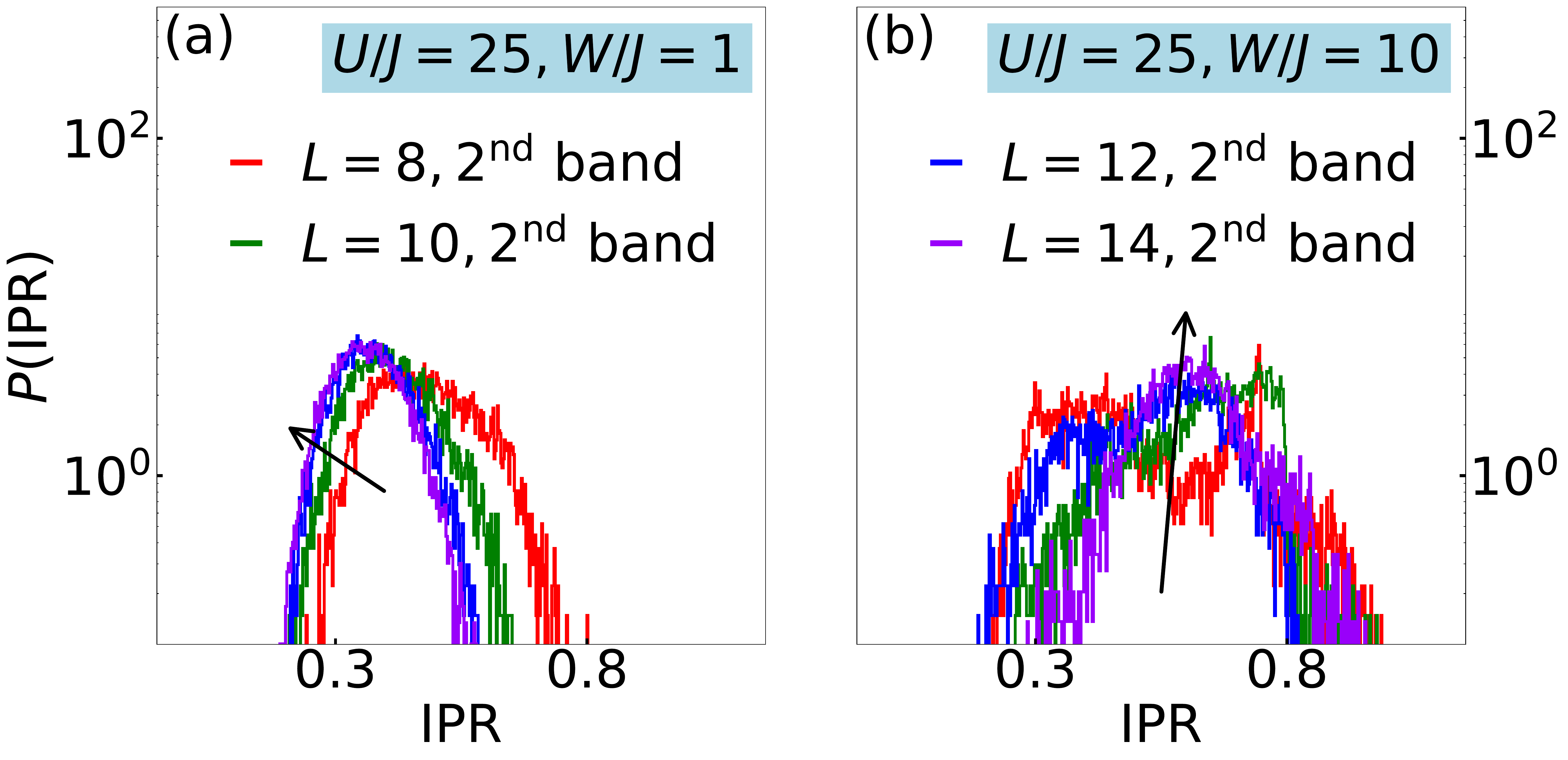}
\caption{{{\it Bose-Hubbard model}: System-size dependence of the full distributions of the IPR
(a) in the ergodic phase and (b) in the MBL phase.
The parameters correspond to the energy densities denoted by the dotted line
in Fig.$~$\ref{entropy_scal}(a), i.e., to the $2^{\rm nd}$ band of the many-body eigenspectra.
The arrows specify increasing system size.}}
\label{IPR_scal}
\end{center}
\end{figure}

Following Ref.$~$\cite{Bera15}, we define the IPR for bosons as
\begin{equation}
{\rm IPR}=\frac{1}{N}\sum_{\alpha=1}^{L} n_{\alpha} \sum_{i=1}^{L} |\phi_{\alpha}^{}(i)|^{4}.
\end{equation}
The IPR measures the real-space localization of the natural orbitals $|\phi_{\alpha}\rangle$.
In Fig.$~$\ref{IPR_scal}, we show the $L$-dependence of the IPR in the second lowest band (for $U/J=25$)
for the same parameters as in Fig.$~$\ref{entropy_scal}. For low disorder $W/J=1$, the IPR distribution has
 a maximum for lower values of IPR with a high-IPR tail which means that the orbitals are mostly delocalized.
The distribution of the IPR exhibits a shift in the position of its maximum towards lower values with increasing system size [see Fig.$~$\ref{IPR_scal}(a)]. 
In the high-disorder regime $W/J=10$, the maxima of the IPR distributions are closer to 
the maximum value of $1$ meaning that the orbitals are mostly localized. Moreover, in the large-disorder regime, the IPR distributions are almost $L$-independent 
[see Fig.$~$\ref{IPR_scal}(b)]. This is consistent with the behavior of the IPR distributions for fermionic systems \cite{Bera15}.

\subsection{Occupations}

\begin{figure}[t!]
\begin{center}
\includegraphics[width=8.5cm]{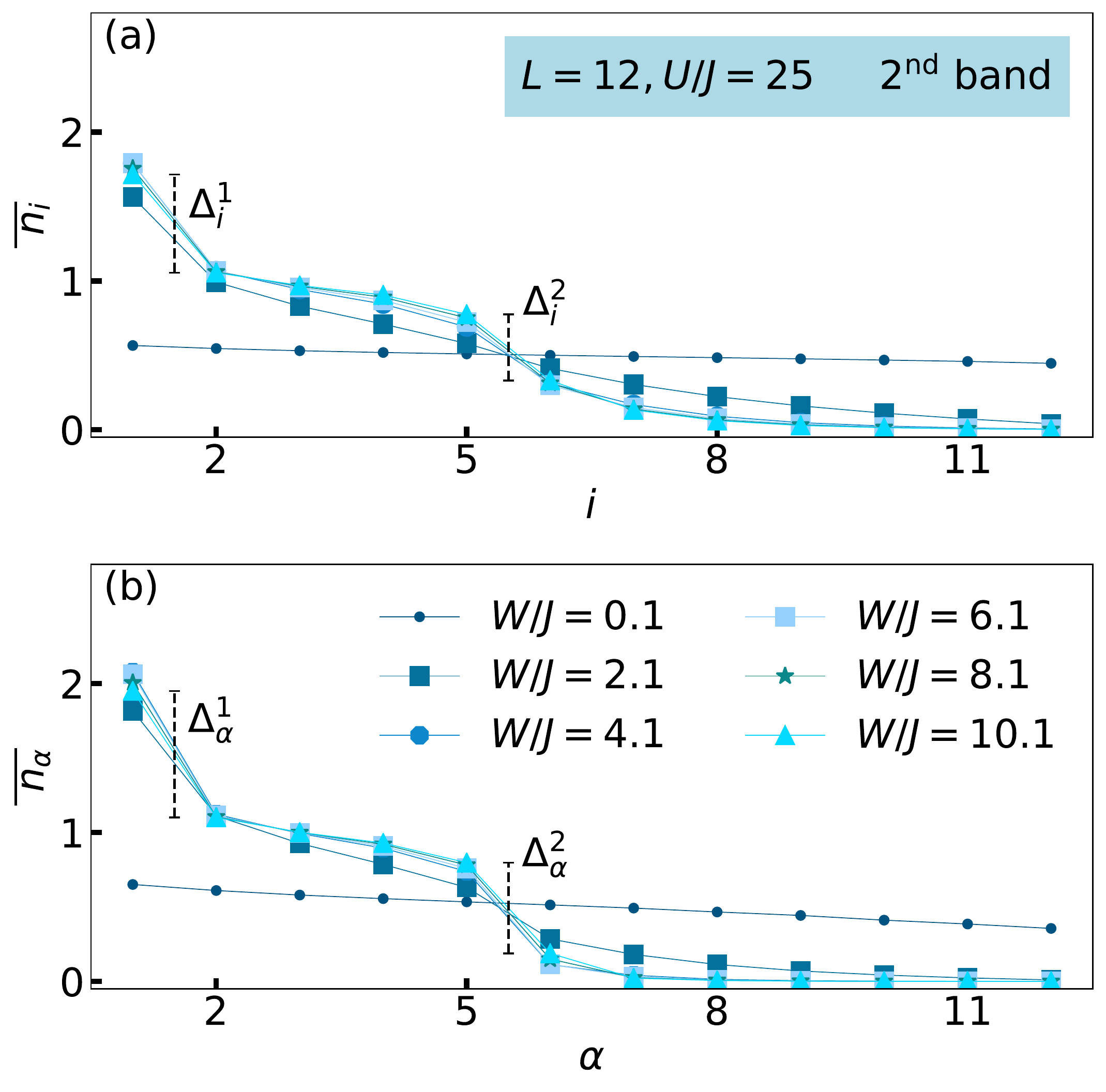}
\caption{{{\it Bose-Hubbard model}: Disorder-averaged and ordered (a) real-space occupations and  (b) occupations of natural orbitals for
the $2^{\rm nd}$ band ($L=12, U/J=25, \epsilon=0.15$).  Both exhibit gaps $\Delta_i^{j}$ and $\Delta^{j}_\alpha$  ($j=1,2$) when first ordered according to $n_1 \geq n_2 \geq n_3 \geq \dots  \geq n_L$
and then averaged over disorder realizations. The vertical dashed lines indicate the location of these discontinuities}}
\label{occpbos}
\end{center}
\end{figure}

\begin{figure}[t!]
\begin{center}
\includegraphics[width=8.5cm]{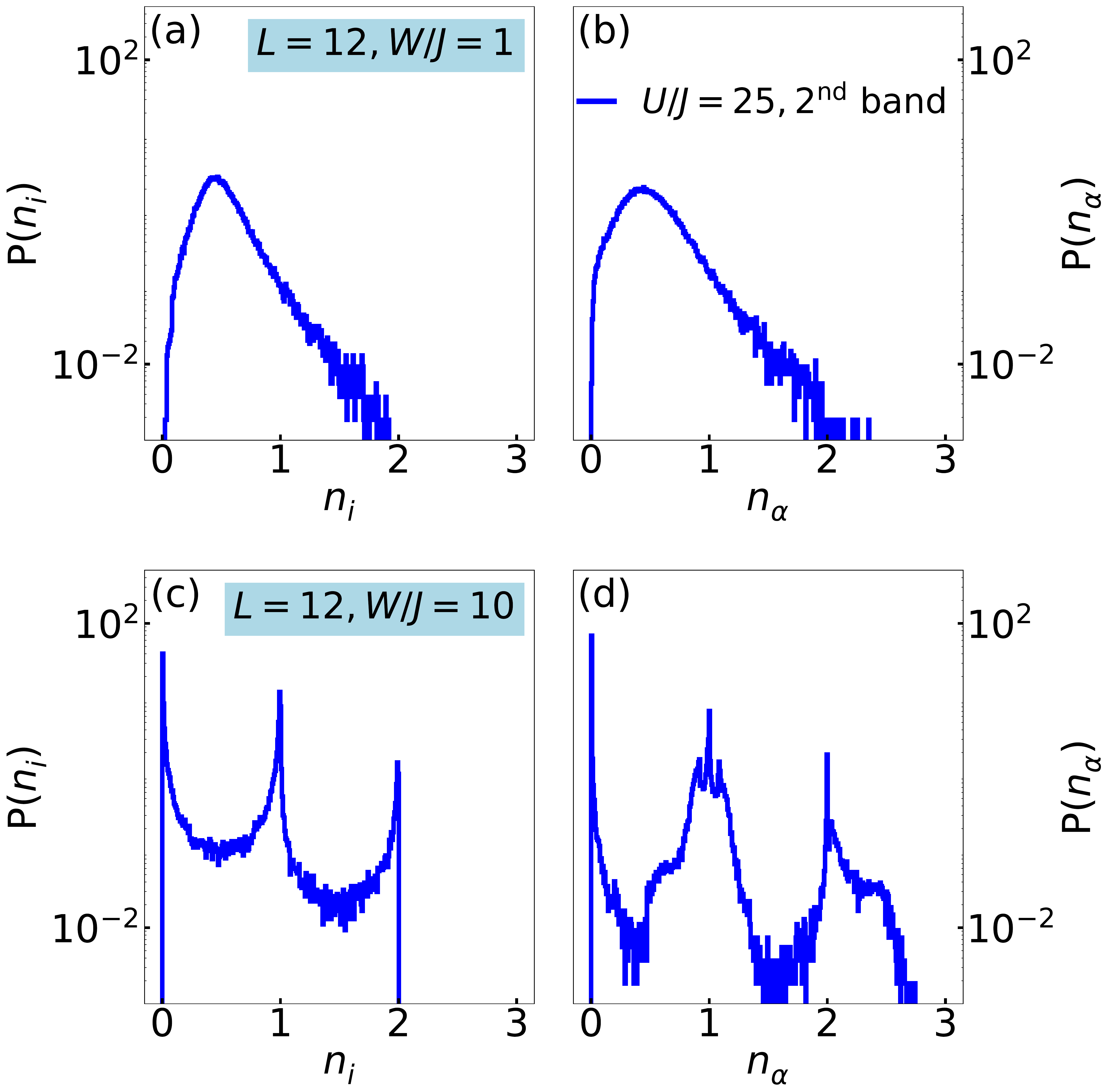}
\caption{{{\it Bose-Hubbard model}: Full distributions of the real-space occupations $n_{i}$ (a) in the
ergodic and (c) in the MBL regime and of occupations of
 natural orbitals $n_{\alpha}$ (b) in the ergodic and (d) 
in the MBL regime for the $2^{\rm nd}$ band ($L=12, U/J=25, \epsilon=0.15$).
}}
\label{occupations}
\end{center}
\end{figure}

In the previous subsection, we have seen  that the natural orbitals contain information about real-space localization. 
In this subsection, we focus on how the occupations, both of the physical sites $|i\rangle$ and of the natural orbitals $|\phi_\alpha\rangle$,
reveal the degree of Fock-space localization.

In analogy to the discussion of  the spin model, we first consider the disorder-averaged occupations.
In Fig.$~$\ref{occpbos}, we show the disorder-averaged occupations for the second band for $L=12$.
For low disorder, the average occupations are a smooth decreasing function. 
For high disorder, we observe that the averaged occupations exhibit a step-like structure where 
the occupations are mostly close to 0,1 or 2. The height of each step between these values is
denoted as gaps $\Delta_i^j$ or $\Delta_\alpha^j$ ($j=1,2$)
[see Fig.$~$\ref{occupations}]. These gaps are analogous to the gaps observed for spins (or hardcore bosons) and fermions.
In the following, we concentrate  on the distributions and the occupation distances as they are better-suited measures
for Fock-space localization.

Examples of the distributions of the occupations $n_i$ and $n_\alpha$ for the second band for $L=12$ 
are displayed in Fig.$~$\ref{occupations}. The first to be noted is that the distributions in the low-disorder regime [see Figs.$~$\ref{occupations}(a) and (b)] are
smooth functions with maxima close to the average density of $0.5$ and with exponentially decaying tails.
In the high-disorder regime, we observe the development of a peak structure. The peaks are located at the integer values $j\in\{0,1,2\}$.
 Higher occupations in the eigenstates are strongly suppressed which is in agreement with the interaction-energy contribution to the energy of the eigenstates in this particular band.
The development of the peak structure in the distributions reflects the ergodic-MBL transition. Thus, analogously to the distributions
of $s_i$ and $s_\alpha$ in the spin system discussed above, the distributions of $n_i$ and $n_\alpha$ indeed reveal the structure
of the Fock-space localization. The distribution of $n_i$ also indicates real-space localization.

\subsection{Quantitative measure of Fock-space localization}

\begin{figure}[b!]
\begin{center}
\includegraphics[width=8.5cm]{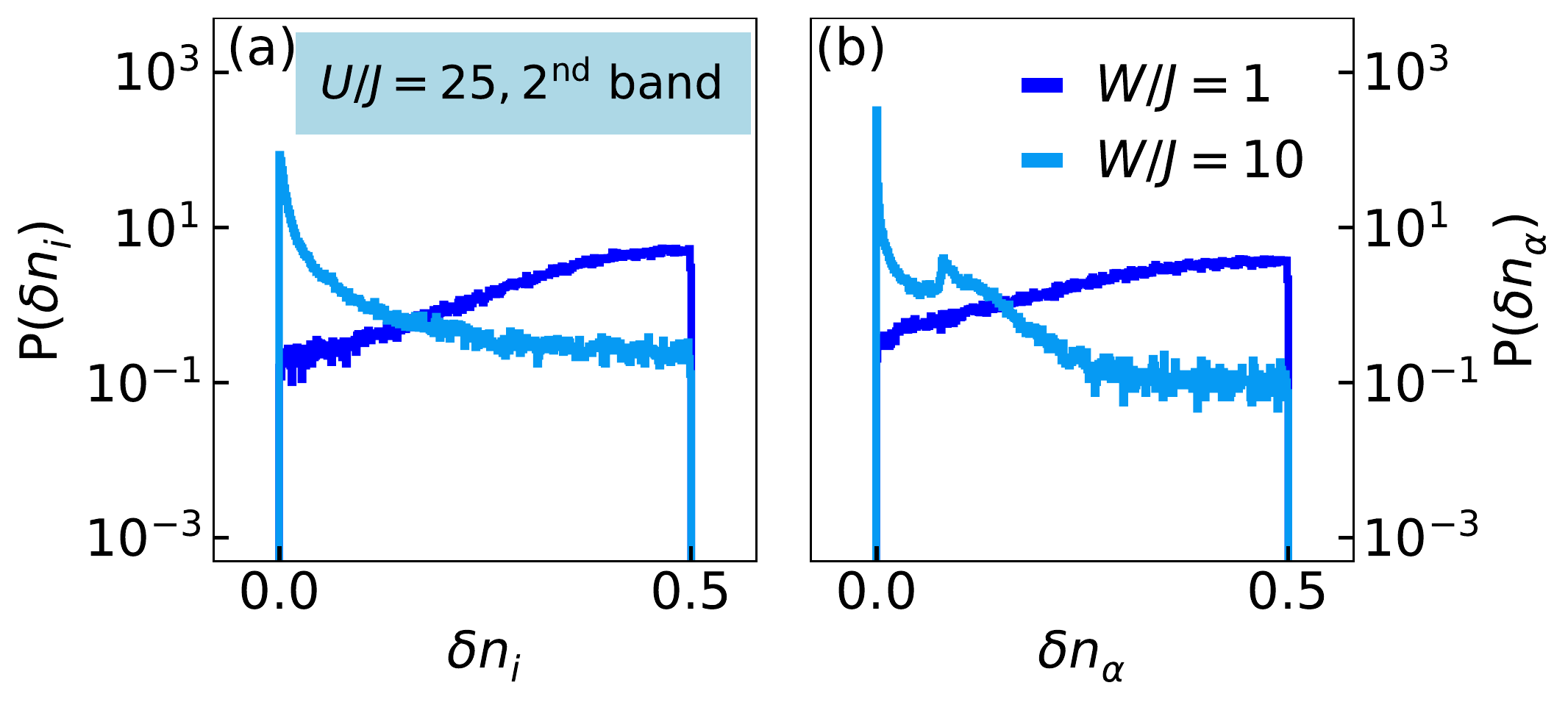}
\caption{{{\it Bose-Hubbard model}: Full distributions of the occupation distances $\delta n_{i}$ of density and  $\delta n_{\alpha}$
of natural-orbital occupations
in the ergodic ($W/J=1$) and in the MBL ($W/J=10$) regimes obtained from the distributions
shown in Fig.$~$\ref{occupations}.}}
\label{distance}
\end{center}
\end{figure}

\begin{figure}[t!]
\begin{center}
\includegraphics[width=8.5cm]{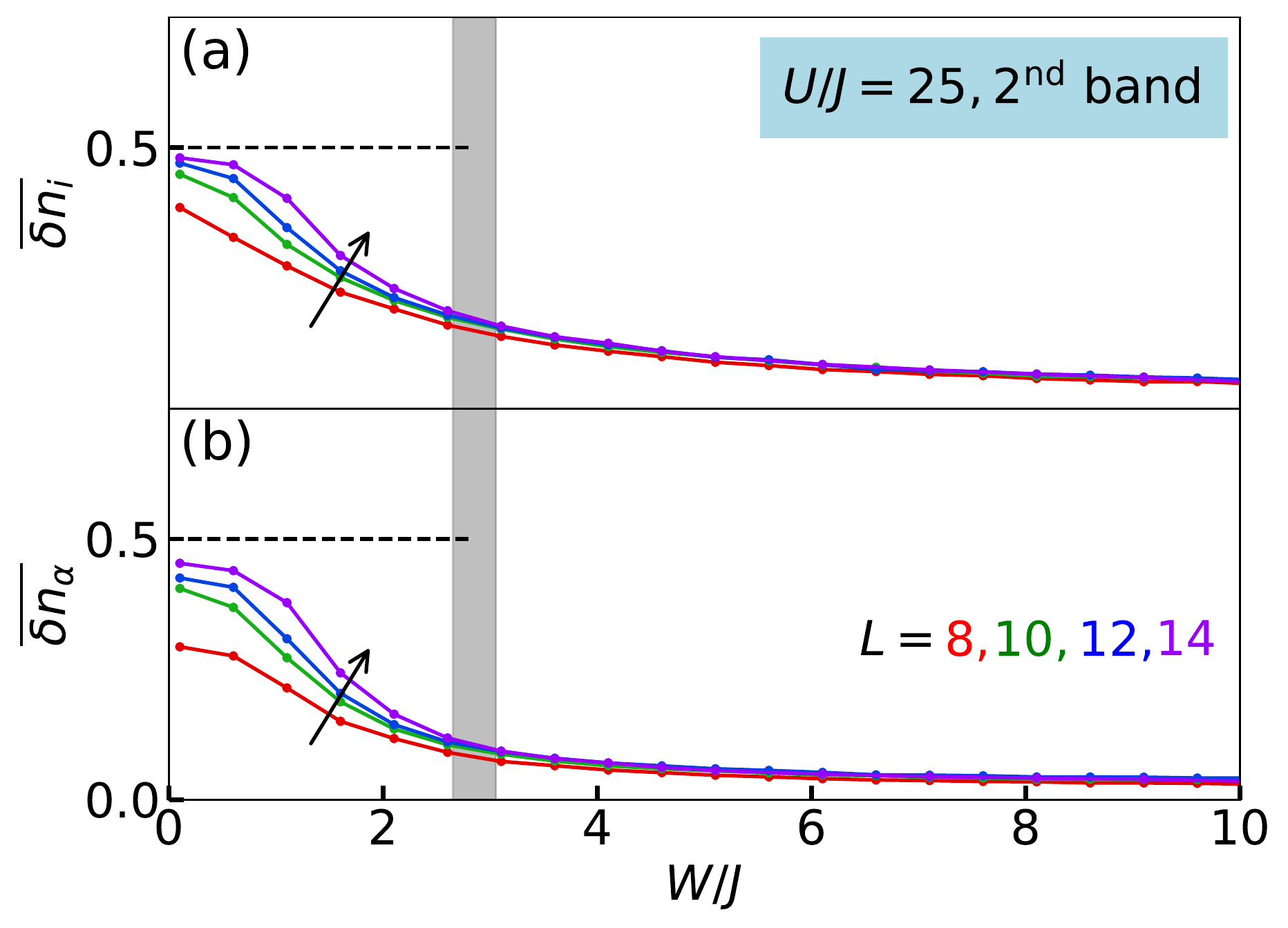}
\caption{{{\it Bose-Hubbard model}: System-size dependence of the average occupation distance
(a) $\overline{\delta n_{i}}$ of densities and (b) $\overline{\delta n_{\alpha}}$
of natural-orbital occupations  a function of $W/J$.
The arrows specify increasing system size.
The parameters correspond to the energy densities denoted by the dotted line
in Fig.$~$\ref{entropy_scal}(a), i.e., to the $2^{\rm nd}$ band of the many-body eigenspectra.
The horizontal dashed line in (a) indicates the filling. The $\overline{\delta n_i}$ is expected to approach this value for $L\rightarrow \infty$
in the ergodic regime. The horizontal dashed line in (b) indicates the upper bound for $\overline{\delta n_\alpha}$.
The vertical line in grey color marks the estimate of the ergodic-MBL transition estimated from visual inspection
of the data in Fig.$~$\ref{scalling4}. }}
\label{scaling3}
\end{center}
\end{figure}
We have seen that the distributions of the site occupations  $n_i$ and the natural-orbital occupations $n_\alpha$ exhibit a 
peak structure in the high-disorder regime which reflects many-body localization. To better quantify the localization, 
we measure, similarly as for the spin system, the distance to the closest integer of the site occupations
\begin{equation}
\delta n_{i}=| n_{i}-[n_{i}] |
\end{equation}
and the distance to the closest integer of the occupations of natural orbitals 
\begin{equation}
\delta n_{\alpha}=|n_{\alpha}-[n_{\alpha}]| \,,
\end{equation}
where $[n_{i}]$ and $[n_{\alpha}]$ are the closest integer to $n_{i}$ and $n_{\alpha}$, respectively.
The results for the distributions of $\delta n_{i}$ and $\delta n_{\alpha}$  for the second band for $L=12$ are displayed in
Fig.$~$\ref{distance} and they show the shift of the maximum of the distribution from $0.5$ to $0$ with increasing
disorder strength. 
Note that for both quantities, $\delta n_i,\delta n_\alpha \leq 0.5$.

In  Fig.$~$\ref{scaling3}, we show  $\overline{\delta n_{i}}$ and 
$\overline{\delta n_{\alpha}}$  as functions of disorder strength $W/J$ for the second band, i.e.,  
for the same parameters as in  Fig.$~$\ref{entropy_scal} and for different system sizes. We observe that the values of $\overline{\delta n_{i}}$ and $\overline{\delta n_{\alpha}}$
are $L$-dependent for the disorder strength $W/J\lesssim 3$ while they are essentially $L$-independent for
$W/J\gtrsim 3$.

\begin{figure}[t!]
\begin{center}
\includegraphics[clip,width=8cm]{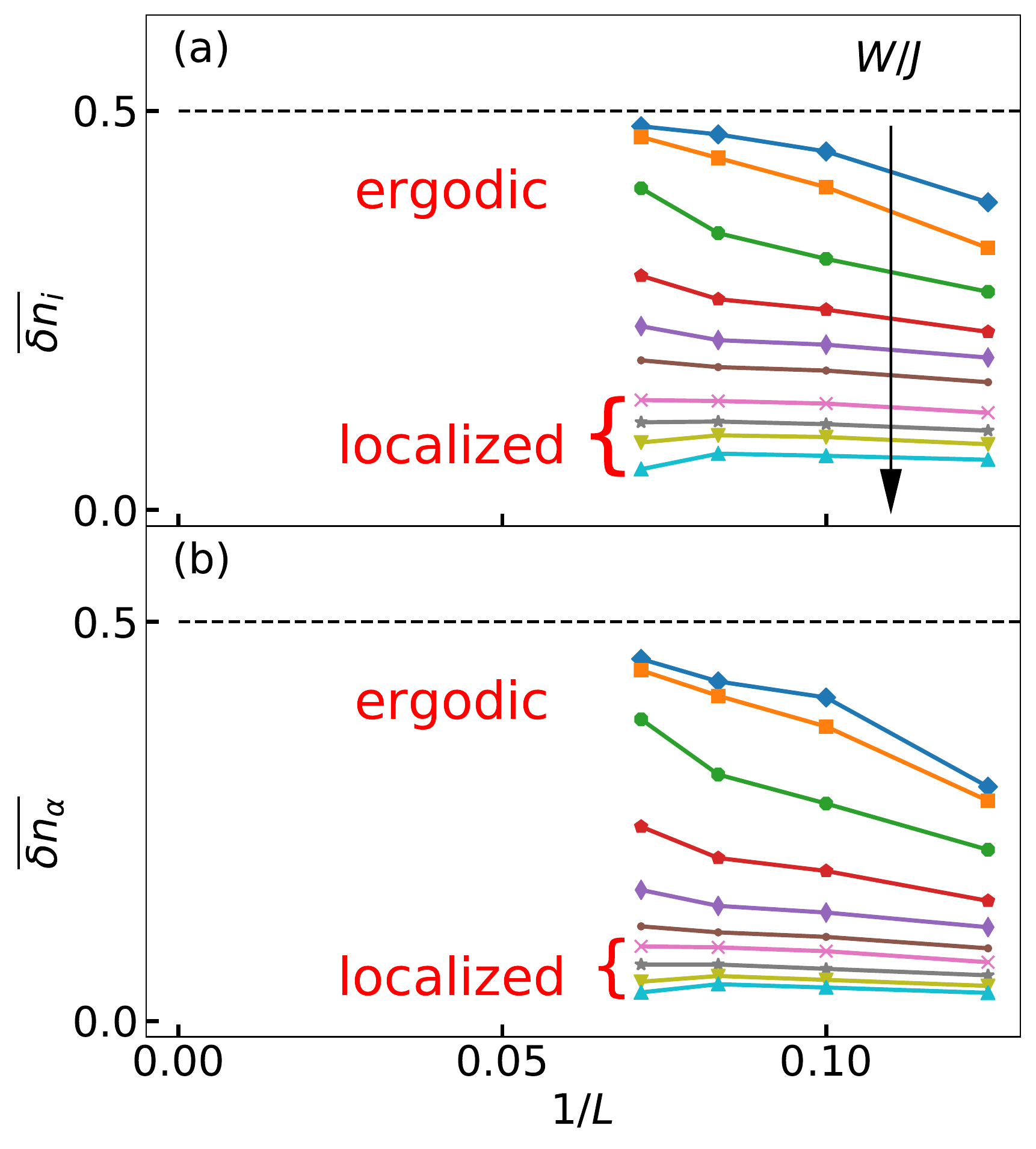}
\caption{{{\it Bose-Hubbard model}:   Average occupation distances (a) $\overline{\delta n_{i}}$ and
(b) $\overline{\delta n_{\alpha}}$  as a function of $1/L$ for $\epsilon=1$ for disorder
strength $W/J=$ 0.1, 0.6, 1.1, 1.6, 2.1, 2.6, 3.1, 4.1, 6.1, 10.1 (various symbols). The arrow specifies increasing disorder strength. The brackets
indicate those data sets that we assign to the localized phase.}}
\label{scalling4}
\end{center}
\end{figure}

To better detect the change of the behavior, we illustrate the $L$-dependences of $\overline{\delta n_{i}}$ and 
$\overline{\delta n_{\alpha}}$ as a function of $1/L$ in Fig.$~$\ref{scalling4}. Clearly, for $W/J\le2.6$, the values of
$\overline{\delta n_{i}}$ and $\overline{\delta n_{\alpha}}$ are increasing functions of $L$ and $\overline{\delta n_i}$ is expected to
approach the upper bound $1/2$ for $L\rightarrow \infty$, consistent with the data. On the other hand, for $W/J\ge3.1$ the values of
$\overline{\delta n_{i}}$ and $\overline{\delta n_{\alpha}}$ appear to saturate
to values much smaller than $1/2$ as a function of $L$. From the visual inspection of the data in Fig.$~$\ref{scalling4}, the behavior changes 
for  $W/J< 3.1$ and we estimate that the transition happens at $ 2.6<W_c/J < 3.1$. This is slightly   higher
than our estimate from the one-parameter scaling of the entanglement entropy of $W_c/J \approx 2.0(1)$. 

The actual values that  $\overline{\delta n_{i}}$ and  $\overline{\delta n_{\alpha}}$  approach in the ergodic phase
clearly depend on filling. For instance, at unit filling, one expects $\overline{\delta n_{i}} \to 0$, while
$\overline{\delta n_{\alpha}}$ is expected to go to a small but energy-dependent value. 
One can introduce a modified occupation distance
\begin{equation}
\tilde \delta_\nu = | n_\nu - n |\,,
\end{equation}
where $\nu=i,\alpha$ and $n$ is the average density or filling.
$\overline{\tilde \delta_i}$ must approach zero in the ergodic phase but remains finite in the MBL phase.
For $\overline{\tilde \delta_\alpha}$, we expect a small but in general nonzero value in the ergodic phase
and a larger limiting value in the MBL phase compared to $\overline{\tilde \delta_i}$.
We have verified this behavior for $n=0.5$  yet observe that the finite-size dependencies of $\overline{\tilde \delta_\nu}$ are larger than
for $\overline{\delta_\nu}$.
 
\begin{figure}[t!]
\begin{center}
\includegraphics[clip,width=9cm]{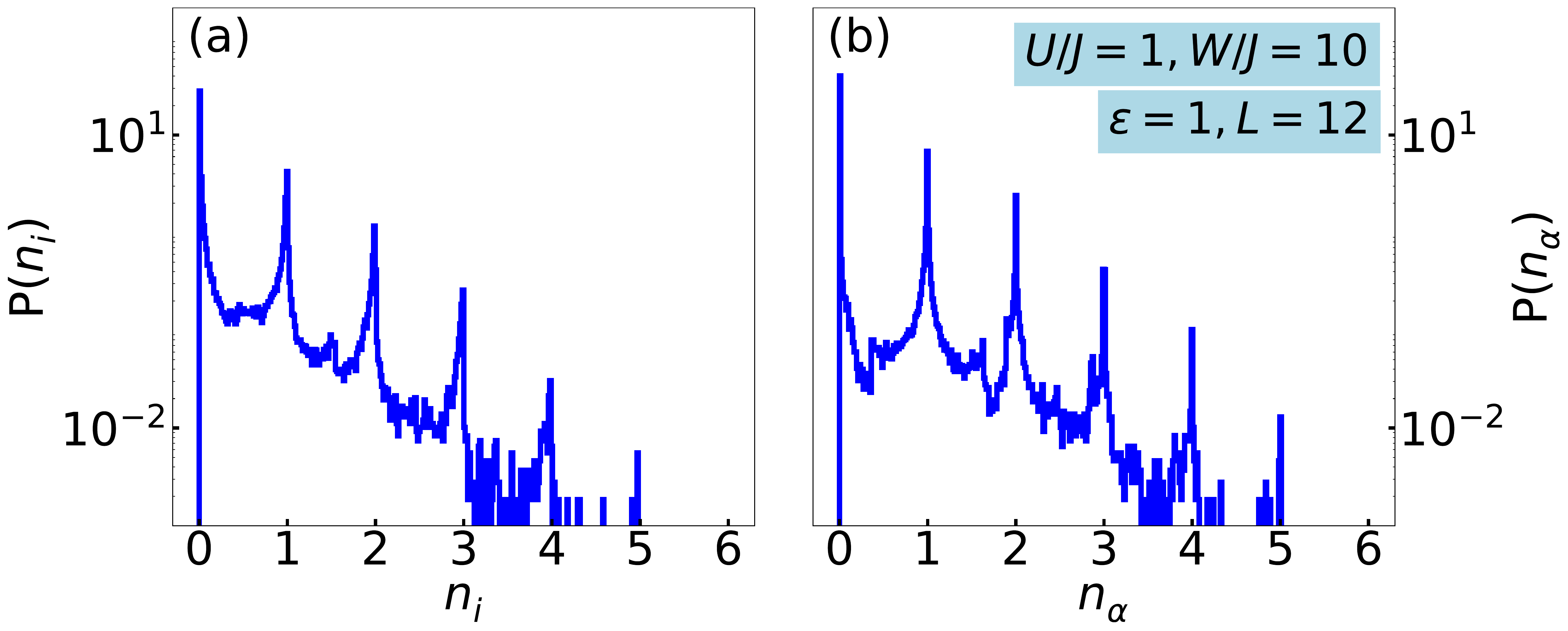}
\caption{{{\it Bose-Hubbard model}: Example for the behavior in the low-interaction regime $U/J=1$. Full distributions of (a) the real-space occupations $n_{i}$, 
and (b) occupations of natural orbitals $n_{\alpha}$, in the MBL regime for the middle of the many-body eigenspectrum ($L=12, \epsilon=1$).}} 
\label{higherocc}
\end{center}
\end{figure}

As a  remark, we mention that the regimes where the occupations can reach values larger than $2$ can be
studied in a similar fashion as the states of the second band. In Fig.$~$\ref{higherocc}, we show an example of the
distributions for $L=12$ and weak interaction strength in the high-disorder regime ($U/J=1, W/J=10)$ for states from the middle of the spectrum.
We observe an analogous peak structure in the distributions of $n_{i}$ and $n_{\alpha}$ with the peaks located around integers $j\in \{1,2,3,4,5\}$ (with
exponentially decreasing weights of the peaks) showing the localization in  Fock space.
The relative weight of the height of the peaks depends on energy density, filling, disorder, and interaction strength.

We conclude that the $L$-dependences of the average occupation distances  $\overline{\delta n_{i}}$ and
$\overline{\delta n_{\alpha}}$ are useful measures for Fock-space localization in the MBL phase.
Analyzing the monotony behavior of the $L$-dependence yields a reasonable estimate for the critical disorder strength,
consistent with other measures.
\subsection{Measuring densities in quantum-gas experiments}

A measurement of $P(n_i)$ should be feasible with quantum-gas microscopes \cite{Schreiber15,Choi16,Kohlert19,Lukin19,Rispoli19}.
In order to obtain  the densities $n_i$ at a certain average density and disorder realization, repeating
projective measurements in the same disorder realization is necessary.
Such experiments with ultracold atomic gases in optical lattices should be capable of reaching much larger system sizes than
exact diagonalization or the shift-and-invert method, which could give better access to the transition.

In principle, there are also other states that can localize particles such as Mott insulators \cite{Bloch08}.
In our case, we work at filling 0.5, where a Mott insulator  would not be realized in the BHM in the absence of a dimerization mechanism. Moreover, one is generally interested
in physics sufficiently high above the ground state in the context of MBL while the Mott insulator  is, strictly speaking, a ground-state
phenomenon. In the Mott insulator at, e.g., unit filling,  the distribution of densities is $P(n_i) \propto \delta(n_i-1)$, which is clearly different from the
behavior in the putative MBL phase [see Fig.~\ref{occupations}(c)]. Obviously, the full characterization of a disordered system should rely on a set of experimental measures, 
including, e.g.,  decay of inhomogeneous density profiles \cite{Schreiber15,Choi16} or density distributions as suggested here. \\

\section{\label{sec:Conclusions}Conclusions}

We  showed that the one-particle density matrix, natural orbitals, and their occupations can be used
to reveal the structure of  real-space and Fock-space localization in systems of interacting disordered bosons. 
The real-space localization is observed in the structure of the natural orbitals, in the system-size dependence of
the inverse participation ratio, and in the full distribution of densities. The Fock-space localization is uncovered via studying 
distributions of occupations and densities.
Particularly, the distributions of the densities $n_i$ and the occupations  of natural orbitals 
$n_\alpha$ are smooth functions in the ergodic regime whereas they develop a peak structure in the MBL regime
where the peaks are at the possible integer eigenvalues of $\hat n_i$ and $\hat n_\alpha$.
Based on this observation, we devised a quantitative measure of localization, the average distance to the 
closest integer of the occupations called occupation distance, and we showed that its system-size dependence 
is strikingly different in the two phases.
This measure can be used to study Fock-space localization for spins, bosons and fermions.

These findings further illustrate the conceptual picture that  many-body
localization  involves localization both in Fock space and in real space. 
An interesting question pertains to a construction of local conserved charges for the MBL phase
of the BHM, i.e., the generalization of l-bits to a system with a large local Hilbert space. 
The distributions of $n_i$ should be accessible in quantum-gas microscope
experiments \cite{Schreiber15,Choi16,Kohlert19,Lukin19,Rispoli19}. It would be interesting to
extend our analysis beyond just the expectation values $n_i$ to a prediction of projective measurements in the MBL phase.\\

\begin{acknowledgements} 
We acknowledge  useful discussions with V. Alba, J. H. Bardarson, I. Bloch, M. Knap, and F. Pollmann. We thank J. Zakrzewski for pointing out Ref.~\cite{Sierant17} to us.
We are indebted to  D. Luitz for sending us  data from Ref.~\cite{Luitz14}.
\end{acknowledgements}

\bibliography{paper_MBL_BHM}
\url

\end{document}